\documentclass[fleqn,usenatbib]{mnras}

\usepackage{newtxtext,newtxmath}

\usepackage[T1]{fontenc}
\usepackage{comment}
\usepackage{subfigure}
\DeclareRobustCommand{\VAN}[3]{#2}
\let\VANthebibliography\thebibliography
\def\thebibliography{\DeclareRobustCommand{\VAN}[3]{##3}\VANthebibliography}

\usepackage{graphicx}	
\usepackage{amsmath}	
\usepackage{multicol}
\usepackage{multirow}
\usepackage{tablefootnote}
\usepackage{verbatim}
\usepackage{array}
\usepackage{soul}
\usepackage{xcolor}
\newcolumntype{P}[1]{>{\centering\arraybackslash}p{#1}}

\defcitealias{Lisiecki2022}{L22}

\usepackage{ulem}

\title[Environments of red nuggets]{Environments of red nuggets at z$\sim$0.7 from the VIPERS survey}

\author[M. Siudek, et al.]{
M. Siudek,$^{1,2}$\thanks{E-mail: msiudek@ifae.es}, K.~Lisiecki$^{3,4}$, J.~Krywult$^{5}$, D.~Donevski$^{3,6}$, C.~P.~Haines$^{7}$, A.~Karska$^{8,4}$,
K.~Małek$^{3,9}$, \newauthor T.~Moutard$^{9}$, A.~Pollo$^{3,10}$
\\
$^{1}$Institut de Física d’Altes Energies (IFAE), The Barcelona Institute of Science and Technology, 08193 Bellaterra (Barcelona), Spain\\
$^{2}$Institute of Space Sciences (ICE, CSIC), Campus UAB, Carrerde Can Magrans, s/n, 08193 Barcelona, Spain\\
$^{3}$National Centre for Nuclear Research, ul. Pasteura 7, 02-093 Warsaw, Poland\\
$^{4}$Institute of Astronomy, Faculty of Physics, Astronomy and Informatics, Nicolaus Copernicus University, ul. Grudziądzka 5, 87-100 Toruń, Poland\\
$^{5}$Institute of Physics, Jan Kochanowski University, ul. Uniwersytecka 7, 25-406, Kielce, Poland\\ 
$^{6}$SISSA, Via Bonomea 265, 34136 Trieste, Italy\\
$^{7}$Instituto de Astronom\'{i}a y Ciencias Planetarias de Atacama (INCT), Universidad de Atacama, Copayapu 485, Copiap\'{o}, Chile\\
$^{8}$Max-Planck-Institut für Radioastronomie, Auf dem Hügel 69, 53121, Bonn, Germany \\
$^{9}$Aix Marseille Univ, CNRS, CNES, LAM, Marseille, France\\
$^{10}$Astronomical Observatory of the Jagiellonian University, ul. Orla 171, 30-244 Krak\'ow, Poland\\
}

\date{Accepted XXX. Received YYY; in original form ZZZ}

\pubyear{2022}

\begin{document}
\label{firstpage}
\pagerange{\pageref{firstpage}--\pageref{lastpage}}
\maketitle

\begin{abstract}
Red ultra-compact massive galaxies, called red nuggets were formed at high redshifts ($\rm{z\sim2-3}$). Survivors of red nuggets, known as relics, observed at lower redshifts ($\rm{z<2}$) are believed to remain almost unchanged since their formation. 
For the first time, we verify the environmental properties of red nuggets at intermediate redshift ($0.5<\rm{z}<0.9$ ) using 42 red, massive ($\rm{log(M_{star}/M_{\odot}) \geq 10.9}$) and ultra-compact ($\rm{R_{e}}<1.5$~kpc) from the VIMOS Public Extragalactic Redshift Survey (VIPERS). 
We found that the increasing fraction of red galaxies, when moving to denser environments, is driven by the red massive normal-size galaxies. Red nuggets, similarly to red intermediate-mass ($\rm{10.4\lesssim log(M_{star}/M_{\odot})<10.9}$) ultra-compact galaxies, are found in various types of environments, with consistent (within $1\sigma$) fractions across all local densities. Analysis of red nugget stellar ages suggests that relics are preferably found in high-density regions while quiescent red nuggets are overabundant in low-density environments. We speculate that red nuggets have survived to lower redshifts via two channels: i) in low-density environments where the fraction of red nuggets decreases as time passes due to (very) limited merger activity, ii) in high-density environments, where the number of red nuggets drops at higher redshift due to merger activity and is preserved at lower redshift as the high velocities of clusters prevent them from being cannibalised. Even more, the fraction of red nuggets in clusters may increase due to the addition of red massive normal-size  galaxies deprived of their envelopes with cosmic time.
\end{abstract}

\begin{keywords}
galaxies: evolution -- galaxies: formation -- galaxies: elliptical and lenticular, cD -- galaxies: clusters: general --  galaxies: structure
\end{keywords}



\section{Introduction}\label{sec:introduction}

Massive, passive, elliptical galaxies, known as early-type galaxies (ETGs,) are perfect laboratories testing the theories of galaxy evolution, as they hold most ($\sim75\%$) of the total stellar mass ($\rm{M_{star}}$) in galaxies at $\rm{z=0}$~\citep[e.g.][]{Fontana2006,Davidzon2017, McLeod2021}. 
Most ETGs have experienced dramatic evolutions in their sizes, morphologies and star formation histories (SFHs) since their formation during a rapid burst at high redshift~\citep[$\rm{z\gtrsim2}$; e.g.][]{Daddi2005,vanDokkum2010,Conselice2011,vanderWel2014}. 
For example, \citealp{vanderWel2014} used 3D-HST and CANDELS surveys to show that the mean effective radii ($\rm{R_e}$) of ETGs at fixed $\rm{M_{star}}$ increase by a factor of 4 from $\rm{z\sim2}$ to $\rm{z\sim0}$. 
Such a drastic change in size over time can be explained by the two-stage hierarchical scenario of galaxy formation and evolution~\citep[][]{Oser2010,Oser2012}. 
In the first phase, at $\rm{z\sim 2-3}$, the cores of massive galaxies are formed through violent dynamical processes~\citep[e.g.][]{Naab2007}, such as early dissipative in-situ star formation~\citep{Keres2005, Dekel2009} and/or gas-rich mergers~\citep{Hopkins2006,Hopkins2008,Wellons2015}. 
The product of this stage -- a red ultra-compact ($\rm{R_{e}<}2$~kpc) massive galaxy, hereafter red UCMG -- is known as a red nugget~\citep{Damjanov2009}. 
The nickname 'red' originates from their physical properties, as they are hosting old stellar populations responsible for their red colours. 
In the second phase, the stellar mass assembly is dominated by dissipationless (also called dry or gas-free) minor  mergers~\citep[e.g.][]{Naab2006,Naab2009}.
These processes cause dramatic size growth accompanied by only a slight increase in the $\rm{M_{star}}$~\citep[e.g.][]{Hilz2013,Tortora2014}. 
However, taking into account the stochastic nature of dry mergers, a small fraction of red nuggets should remain undisturbed without accreting any stars through cosmic time~\citep[i.e. from simulations the fraction of survivals is 1--10\%,   ][]{Hopkins2009,Quilis2013}.  
Those survivors of high-redshift red nuggets observed in the local Universe  characterised by the stellar ages as old as the Universe are known as relics and were shown to be $\sim4$ times more compact than massive galaxies observed today~\citep{Daddi2005, Trujillo2007}. 

Many studies have already shown that the SFHs of ETGs depend strongly on both their $\rm{M_{star}}$ and the environment in which they reside. 
Observationally, the majority of ETGs are located in dense environments, such as groups and clusters, whereas most of active spiral galaxies are found in less dense areas~\citep[e.g.][]{Dressler1980,Peng2010,Guglielmo2015,Siudek2022}.   
Such morphological segregation of galaxies appears to be a universal characteristic of galaxy populations~\citep[e.g.][]{Balogh1997, Lewis2002,Hogg2003,Kauffmann2004} up to at least $\rm{z\sim 2}$~\citep[e.g.][]{Cooper2007,Dressler1997,Tasca2009,Chuter2011,  Andreon2020, Sazonova2020, Siudek2022}. 
The puzzling role of internal and external processes, known as the nature vs nurture dilemma, is still under debate~\citep[e.g.][]{DiMatteo2005,Peng2010,PaulinoAfonso2019,Tortora2020,Turner2021,Siudek2022, Siudek2023}. 
It seems that the galaxy's evolution has been driven by a mixture of both factors, but how the properties of ETGs change when moving to denser environments remains unclear.  


The past merger history should be reflected in the size of ETGs: in high-density (HD) environments, massive galaxies formed early and their sizes should grow rapidly through dry merging, while in low-density (LD) environments, where merger activity is less frequent, ETGs should have relatively small sizes for a given mass.
Such a correlation with the environment has been found at $\rm{1<z<2}$~\citep{Papovich2012,Lani2013}, but at intermediate redshift ($\rm{0.2<z<1}$) studies are not conclusive: some groups have found that ETGs are larger when moving to denser environments~\citep[e.g.][Donevski et al. in prep]{Capak2007,Cooper2012,Siudek2022} and others have not found such a correlation~\citep{HuertasCompany2013,Kelkar2015}.  
Similarly, in the local Universe, some groups found a positive correlation between the size of ETGs and the environment~\citep{Yoon2017}, whereas the others found no such dependence~\citep{Maltby2010,Nair2010,Cappellari2013,HuertasCompany2013}. 
 

Intuitively, we would expect to find red nuggets in less hostile, LD environments. 
Using the Illustris simulations, \cite{Wellons2016} predicted that massive compact galaxies should be predominantly found in under-dense environments and in satellites which formed in larger groups at low redshifts. 
However, the first observed UCMGs have been found to be located in clusters~\citep[][]{Valentinuzzi2010b,Valentinuzzi2010a,poggianti2013}. 
Their presence in clusters has been justified by the high-velocity dispersions of clusters that should prevent UCMGs from undergoing mergers, allowing them to survive until today~\citep{poggianti2013}. 
Yet, it is unclear whether UCMGs residing in clusters are survivors of red nuggets or the cores of larger galaxies deprived of their outer stellar populations~\citep[e.g.][]{Buitrago2018,Tortora2020, Flores2022}.
Several different studies have demonstrated that UCMGs are not limited to a single environment, and instead can be found in a similar range of environments to their normal-size counterparts~\citep{Damjanov2015,Buitrago2018,Tortora2020}. 
The discrepancies between the environmental properties of red nuggets between studies can be at least partially attributed to different methods of the estimation of galaxy properties and/or discrepancies in the definition of environmental quantities in all these studies. 
Moreover, selection effects in the studied galaxy samples may also have an impact on the final conclusions. 
However, these discrepancies demonstrate that we are far from getting a clear picture of the impact of the environment on the sizes of red nuggets or, more generally, of ETGs.  

Recently, different works have contributed to the census of UCMGs at $\rm{z<0.6}$~\citep[][and references therein]{Scognamiglio2020,Spiniello2021a, Spiniello2021b}. 
However, those systems are very rare. 
The largest spectroscopic sample of UCMGs
contains only 92 objects with $\rm{log(M_{star}/M_{\odot}) \geq 10.9}$ and $\rm{R_{e} < 1.5}$~kpc, at $\rm{0.1<z<0.5}$~\citep{Tortora2018,Scognamiglio2020},  leaving the intermediate redshift range ($\rm{0.5<z<1.0}$) still unexplored.

In~\cite{Lisiecki2022}, hereafter \citetalias{Lisiecki2022}, we described the first sample of 77 red nuggets at intermediate redshift,  
$\rm{0.5<z<1.0}$, drawn from the VIMOS Public Extragalactic Redshift Survey~\citep[VIPERS;][]{Scodeggio2018}. 
This catalogue of red nuggets stands out from the others with: 
i) the coverage of a unique redshift range, ii) the selection based on the large-area, high-completeness spectroscopic survey, and iii) the wealth of auxiliary VIPERS data, including environmental properties.
VIPERS' red nugget sample provides the next step to understanding the evolution of ETGs.  
Thanks to its combination of large volume coverage and sampling (comparable to the 2dF Galaxy Redshift Survey in the local Universe) and the wealth of ancillary information, VIPERS provides a unique opportunity for environmental studies at intermediate redshifts.  
Thus, for the first time, we can examine whether red nuggets at intermediate redshifts prefer dense environments, as the first observations suggest, or are found among different environments as more recent studies advocate.

This paper is structured as follows: 
in Sec. \ref{sec:data} we present the VIPERS data. 
Section~\ref{sec:sample_selection} explains the sample selection and describes the red nugget catalogue. 
In Sec.~\ref{sec:environment_properties}, we present the environmental properties of red nuggets. 
In Sec.~\ref{sec:comparison_literature} we compare our findings with the literature, and discuss red nugget environmental properties in Sec.~\ref{sec:discussion}. 
The summary is presented in Sec.~\ref{sec:summary}. 
Throughout this paper the cosmological framework with $\Omega_{m}$ = 0.30, $\Omega_{\Lambda}$ = 0.70, and $H_{0}=70$ $\rm{km s^{-1} Mpc^{-1}}$ is assumed.

\section{Data}\label{sec:data}
We retrieved our sample from VIPERS~\citep[][]{Scodeggio2018}. 
VIPERS is a completed European South Observatory (ESO) Large Program that mapped 86\,775 galaxies at redshifts $0.5<z<1.2$ with a wavelength coverage of 5\,500--9\,500\,{\AA} and a resolution of $R\sim220$. 
VIPERS maps an area of $\sim$23.5\,$\deg^2$ within the W1 and W4 fields of the Canada-France-Hawaii Telescope Legacy Survey Wide (CFHTLS-Wide) to a limit of $\rm{i< 22.5}$ mag. 
A~detailed description of the survey is given by~\cite{Guzzo2014} and ~\cite{Scodeggio2018}. 
The specification of the data reduction pipeline with the redshift quality system is described by~\cite{Garilli2014}.	

\subsection{Physical properties}\label{sec:physical_properties}

The physical properties of the VIPERS sample were derived from spectral energy distribution (SED) fitting based on the $\rm{u}$, $\rm{g}$, $\rm{r}$, $\rm{i}$, $\rm{z}$ fluxes from the CFHTLS T0007 release~\citep{goranowa09}, $\rm{FUV}$ and $\rm{NUV}$ measurements from GALEX~\citep{Martin2005}, near-infrared  $\rm{K_{s}}$ band from WIRCAM~\citep{Puget2004S} complemented by $\rm{K_{vista}}$ from the VIDEO VISTA survey~\citep{jarvis13}. 
In addition, we used mid-infrared observations from the Spitzer WIDE-area Infrared Extragalactic Survey (SWIRE) of the XMM-Large-Scale Structure~\citep[XMM-LSS,][]{Pierre2004} field, which overlaps with the W1 field. 
We incorporated also photometry from the NASA Wide-field Infrared Survey Explorer~\citep[WISE,][]{Wright2010}. 
Since the $\rm{M_{star}}$ used in~\citetalias{Lisiecki2022} were derived only for a sample of 6\,961 UCMG candidates, we performed the SED fitting for VIPERS galaxies used in this paper (see Sec.~\ref{sec:data}), to ensure that the same prescription is used to derive physical properties among all analysed galaxies. 

The $\rm {M_{star}}$ and star formation rates (SFRs) among other physical properties, were derived with a state-of-the-art SED-fitting tool, the Code Investigating GALaxy Emission~\citep[CIGALE version 2022.0;][]{Boquien2019}. 
This tool is based on the principle of the energy balance between dust-absorbed stellar emission in the ultraviolet and optical bands and its re-emission in the infrared. 
CIGALE has been thoroughly tested on VIPERS photometry to estimate the physical properties of star-forming galaxies~\citep{Figueira2022,Pistis2022}, galaxies hosting active galactic nuclei~\citep{Vietri2021, Siudek2023, Mezcua2023}, passive galaxies~\citep{Lisiecki2022}, or the whole population~\citep{Turner2021}. 
In this work, following the methodology introduced in~\citetalias{Lisiecki2022}, we used a delayed SFH model and ~\cite{Bruzal2003} single stellar population models assuming a~\cite{Chabrier2003} initial mass function and solar metallicity to build a grid of models. 
We also utilised the standard nebular emission model from~\cite{Inoue2011} and modelled dust attenuation using the  \citealp{Calzetti2000} attenuation curve. 
The reprocessed dust emission was modelled using the dust emission models of~\cite{dale2014}. 
SEDs of these models were then fitted to the VIPERS galaxy SEDs using a Bayesian-like analysis, where the quality of the fit is expressed by the reduced $\chi^2$. 
The estimates and errors of physical properties are the likelihood-weighted mean and standard deviation of all the models, respectively~\citep{Boquien2019}. 
The detailed description of the SED-fitting procedure for the VIPERS galaxies used in the following analysis can be found in~\citetalias{Lisiecki2022}.

\subsection{Structural parameters}\label{sec:morphology}

The structural parameters were derived by fitting point spread function (PSF)-convolved S\'ersic profiles to the observed $\rm{i}$-band CFHTLS-Wide images~\citep{Krywult2017}.  
In short, the galaxy was first detected in each image with SExtractor~\citep{Bertin1996} and then the S\'ersic model was convolved with a local PSF using GALFIT~\citep{Peng2002,Peng2010}. 
The best fit was found based on the Levenberg-Marquardt algorithm to minimise the goodness, $\chi^2$, of the fit~\citep{Peng2002}. 
GALFIT provides the semi-major axis ($\rm{a_e}$), the axial ratio ($\rm{b/a}$) of the profile, from which the circularised effective radius ($\rm{R_e=a_e\sqrt{b/a}}$) was derived, as well as the S\'ersic index, $n$. 
\cite{Krywult2017} have conducted several tests to prove the robustness of derived structural parameters. 
In particular, the authors have simulated galaxies using GALFIT and added them into the CFHTLS images. 
The results of this test returned the errors of effective radii measurements at the level of 4.4$\%$ (12$\%$) for 68$\%$ (95$\%$) of VIPERS sample. 
Hence, it seems that the automatic estimations of structural parameters could be trusted in preselecting VIPERS galaxies based on their shapes and investigating their change with the environment. 
A~detailed description of the structural parameters can be found in~\cite{Krywult2017}. 
In this work, we further verify the accuracy of the derived $\rm{R_e}$ of ultra-compact galaxies in different environments. In App.~\ref{app:size_validation} using Monte Carlo simulations, we show that PSF decomposition significantly improved the accuracy of recreating the proper radii of compact galaxies (see Fig.~\ref{fig:re_simulations}). 

\subsection{Local densities}\label{sec:environment}

The local environment was characterised by the density contrast, $\rm{\delta}$, smoothed over a cylinder with a radius equal to the distance of the fifth nearest neighbour, $\rm{D_{p,5}}$ \citep{Cucciati2017}, and defined as: 
\begin{equation}
 \delta(\rm{RA},\rm{DEC},z)=[\rho(\rm{RA},\rm{DEC},z)-\langle\rho(z)\rangle]/\langle\rho(z)\rangle,
\end{equation}
where $\rho(RA,DEC,z)$ is the local density at the comoving position of each galaxy and $\langle\rho(z)\rangle$ is the mean density at that redshift. 
The mean density is estimated using all galaxies that trace the density field (tracers) within a cylindrical top-hat filter with a half-length of $\pm1000$ km/s and the radius corresponding to the $\rm{D_{p,5}}$. 
The tracers are selected from a volume-limited sample that includes both spectroscopic and photometric galaxies with a cut $\rm{M_{B} \le (-20.4 - z)}$, providing complete tracer samples up to $\rm{z =0.9}$. 
This tracer sample provides a comoving number density that does not evolve, therefore is not affected by discreteness effects that change with redshift. 
Extending the redshift out to $\rm{z=1}$ (the upper limit of VIPERS red nugget catalogue) by using even brighter tracers, would increase the $\rm{D_{p,5}}$ at fixed $\rm{\delta}$ at the price of computing the density field on much larger scales. 
A~detailed description of the environmental properties can be found in~\cite{Cucciati2014, Cucciati2017}.

\section{Sample selection}\label{sec:sample_selection}

In the following Section, we describe the samples used in this paper. 
We built a sample of red nuggets from the first red nugget catalogue at intermediate redshift presented in~\citetalias{Lisiecki2022}. 
In addition, we created four control samples to trace {whether} the environmental properties of red nuggets are different with respect to red galaxies.

The catalogue presented in~\citetalias{Lisiecki2022} is based on 36\,157 VIPERS galaxies with secure redshift measurements~\citep[provided by a confidence level higher than 95\%,][]{Garilli2014} and reliable $\rm{R_e}$ estimations~\citep[provided by 
$\rm{R_e}>3\,\sigma_{R_e}$, S\'ersic index, $n>0.2$, and quality of the fit, $\rm{\chi^2<1.2}$, see details in][]{Krywult2017} observed at $\rm{0.5<z<1.0}$.  
In this work, we further restricted this initial sample to the 28\,045 galaxies with reliable $\delta$ estimations~\citep[i.e. for which at least 60\% of the cylinder volume overlaps with the VIPERS observed volume, ][]{Cucciati2017}. 
Moreover, we required the sample to be complete in stellar mass, i.e. with $\rm{M_{star}}$ above a mass limit for passive and active galaxies as defined in~\cite{Davidzon2016}. 
This limit corresponds to $\rm{log(M_{star}/M_{\odot})=}$ 10.39 (10.18), 10.65 (10.47), 10.86 (10.66) for redshift bins $\rm{0.50<z\leq0.65}$, $\rm{0.65<z\leq0.85}$, $\rm{0.80<z\leq0.90}$ for passive (active) galaxies, respectively. 
Passive and active galaxies were identified based on the NUV-r, vs r-K, diagram, called NUVrK diagram ~\cite[][see also App.~\ref{app:colours}]{Arnouts2013,moutard2016b}. 
The application of $\rm{M_{star}}$ cuts to passive and active galaxies limited our sample to 7\,306 galaxies, hereafter the \textit{VIPERS sample}. 
This sample is used in the next steps to select a sample of red nuggets and generate control samples presented in Sec.~\ref{sec:rednugget_sample} and \ref{sec:control_sample}, respectively.

Each galaxy from the \textit{VIPERS sample} is provided with a statistical selection weight $\rm{w}$ accounting for survey incompleteness.  
The selection weight reflects the {representativeness} of a given galaxy with respect to the underlying parent photometric catalogue. 
The selection weight, $\rm{w}=1/(\rm{TSR}\times\rm{SSR}\times\rm{CSR})$, takes into account three selection functions: the target sampling rate (TSR), the spectroscopic success rate (SSR), and the colour sampling rate (CSR). 
The overall sampling rate, i.e. TSR $\times$ SSR $\times$ CSR is on average 45\%. 
Further details about these selection functions are provided in~\citealp{Garilli2014} and~\citealp{Scodeggio2018}. 
To account for mass incompleteness introduced by Malmquist bias, each VIPERS galaxy is weighted also by the fraction of the survey volume in which the galaxy would be still observable (using minimal and maximal redshifts at which it could be observed), following the  $\rm{1/V_{max}}$ method~\citep{Schmidt1968}. 
In this work, the number of galaxies is weighted according to the selection weight and $\rm{1/V_{max}}$ method. 

\subsection{Red nugget sample}\label{sec:rednugget_sample}

In~\citetalias{Lisiecki2022}, the sample of 77 red nuggets was created by selecting red, massive and ultra-compact galaxies. 
As the number of candidates may change by orders of magnitude depending on the applied definition, VIPERS red nuggets were selected using the most restrictive criterion found in the literature~\citep{Tortora2016,Scognamiglio2020}, i.e. defined as those with a stellar mass $\rm{log(M_{star}/M_{\odot}) \geq 10.9}$ $\rm{M_{\odot}}$ and effective radii $\rm{R_e}<$ 1.5 kpc. 
Furthermore, we restricted our sample to only red objects (based on the NUVrK diagram) without emission lines to select only galaxies without star formation activity{. This ensures that all VIPERS red nuggets have low SFRs ($\rm{SFR<1}$ $\rm{M_{\odot}yr^{-1}}$)}. 
We expect to find some relics among these. Relics are red nuggets with stellar populations as old as the Universe~\citep[ e.g.][]{Buitrago2018}. 
Based on the strength of the 4000\AA~break~\citep[D4000,][]{Balogh1999}, a spectral feature used to characterise the stellar age~\citep[e.g.][]{Kauffmann2003,Haines2017,siudek2017,Renard2022}, we showed that VIPERS red nuggets are populated on average by old stellar populations (see details in~\citetalias{Lisiecki2022}). 
{In Sec.~\ref{sec:rednugget_ages} we further explore the stellar ages of the red nuggets.} 
In this paper, we are focusing on the environmental characteristics of red nuggets.  
Thus, we limited the sample of \citetalias{Lisiecki2022} to only those with reliable environment estimations \citep[i.e. for which at least 60\% of the cylinder volume overlaps with the VIPERS observed volume,][]{Cucciati2017}. 
This selection removed 35 (45\%) of the initial 77 red nuggets, defining a sample of 42 red nuggets, hereafter the \textit{red nugget sample}.

\subsubsection{Validation of the \textit{red nugget} sample}\label{sec:rednugget_validaiton}

\begin{figure*}
	\centerline{\includegraphics[width=0.99\textwidth]{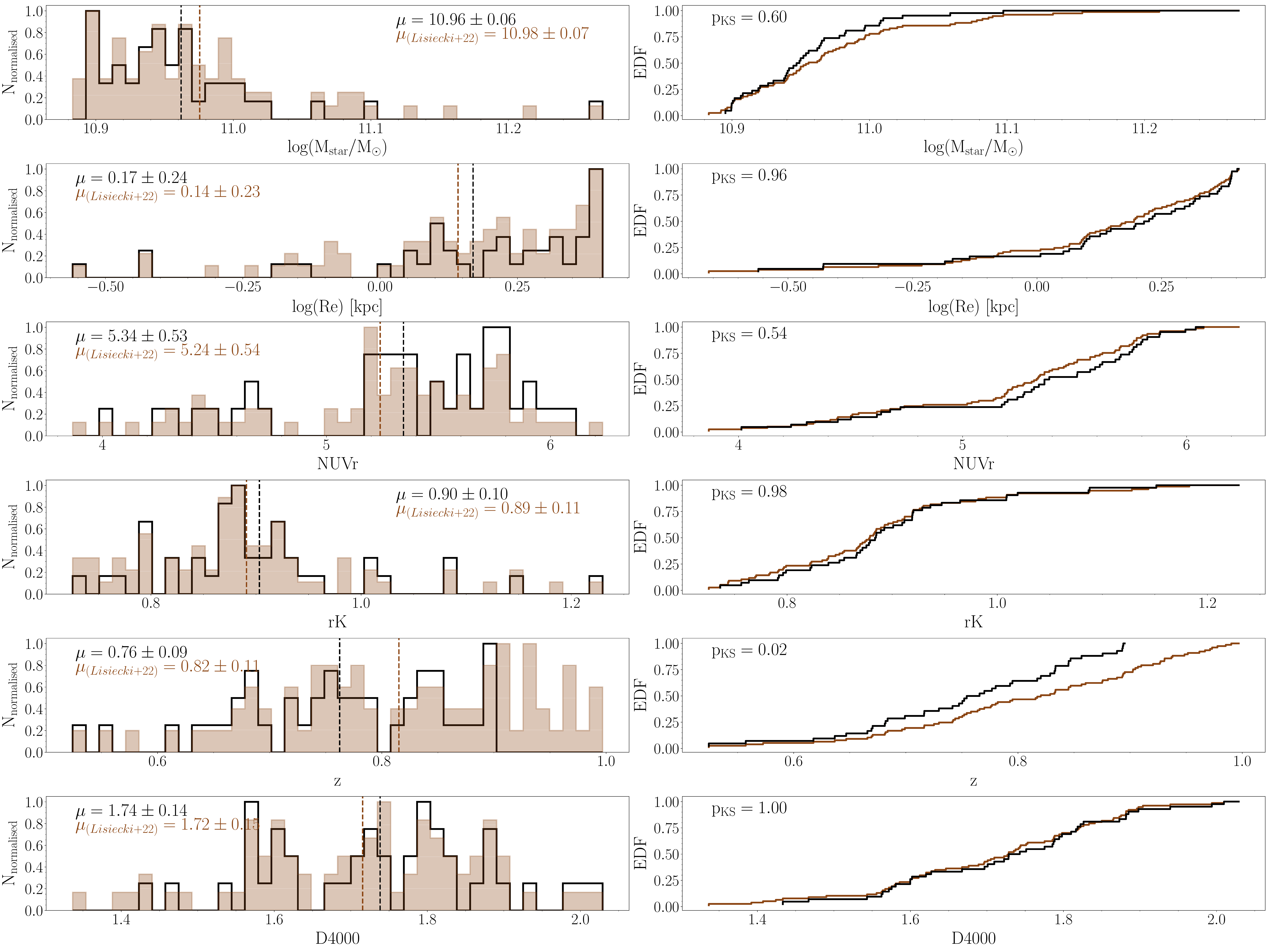}}
	\caption{Left panels: Distributions of $\rm{M_{star}}$, $\rm{R_e}$, NUV-r, r-K, z, D4000 of \textit{red nugget sample} (42 galaxies, in black) and complete red nugget sample (77 galaxies) from~\citetalias{Lisiecki2022} (in brown). The mean values and associated standard deviations are reported in the plots. Right panels: The same as for the left panel but for the empirical distribution functions (EDFs). The KS tests are reported in the plots. }
	\label{fig:rednugget_validatio}
\end{figure*}

The sample used in this paper is built of 42 out of 77 VIPERS red nuggets from~\citetalias{Lisiecki2022}. Thus, we verify whether the properties of the \textit{red nugget sample} are biased with respect to the original sample of red nuggets. We examine whether the distributions of $\rm{M_{star}}$, $\rm{R_e}$, colours (NUV-r and r-K), redshift and D4000 change when selecting our sample out of the original one. As shown in Fig.~\ref{fig:rednugget_validatio} the main properties, except redshift, of the \textit{red nugget sample} are similar to the ones for the original sample as indicated by their mean values and Kolmogorov–Smirnov (KS) tests. The KS test is a commonly used non-parametric test based on the distance between two cumulative distributions. 
High p-values of the KS test ($\rm{p_{KS}>0.5}$) are found when comparing the distributions of $\rm{M_{star}}$, $\rm{R_e}$, NUV-r, r-K, D4000 of the two samples: the \textit{red nugget sample} and the original sample, indicating that the hypothesis that the two distributions originate from the same parent distribution cannot be rejected.
The only difference in the samples is seen in the redshift distribution (see the fifth panel in Fig.~\ref{fig:rednugget_validatio}) as the original sample extends to $\rm{z=1.0}$, while the  environmental studies can be performed only on galaxies observed out to $\rm{z=0.9}$. 
When considering the same redshift range for both samples (i.e. $\rm{z\leq0.9)}$ the redshift distributions are indistinguishable as confirmed by high p-values of the KS test ($\rm{p_{KS}=1}$). Thus, the different redshift distributions should not affect our studies. 

\subsubsection{{Stellar ages of the \textit{red nugget} sample}}\label{sec:rednugget_ages}

{As already mentioned in Sec.~\ref{sec:introduction}, red nuggets that have survived to the present-day Universe avoiding merger events, nicknamed as \textit{relics}, are characterised by stellar ages as old as the Universe (i.e with stellar ages $\geq 10$ Gyr in galaxies observed at redshift $\rm{z\sim}0$). The low-end of the D4000 distribution for the VIPERS \textit{red nugget} sample (see the last panel in Fig.~\ref{fig:rednugget_validatio}) is already below the threshold found for red passive galaxies~\citep[D4000>1.5,][]{Kauffmann2003} suggesting that our \textit{red nugget} sample includes galaxies with more recent star formation activity. Thus, we validate whether \textit{red nuggets} observed at $\rm{z\sim0.7}$ represents the relic population or rather galaxies with signatures of more recent star formation activity.} 

\begin{figure*}
	\centerline{\includegraphics[width=0.99\textwidth]{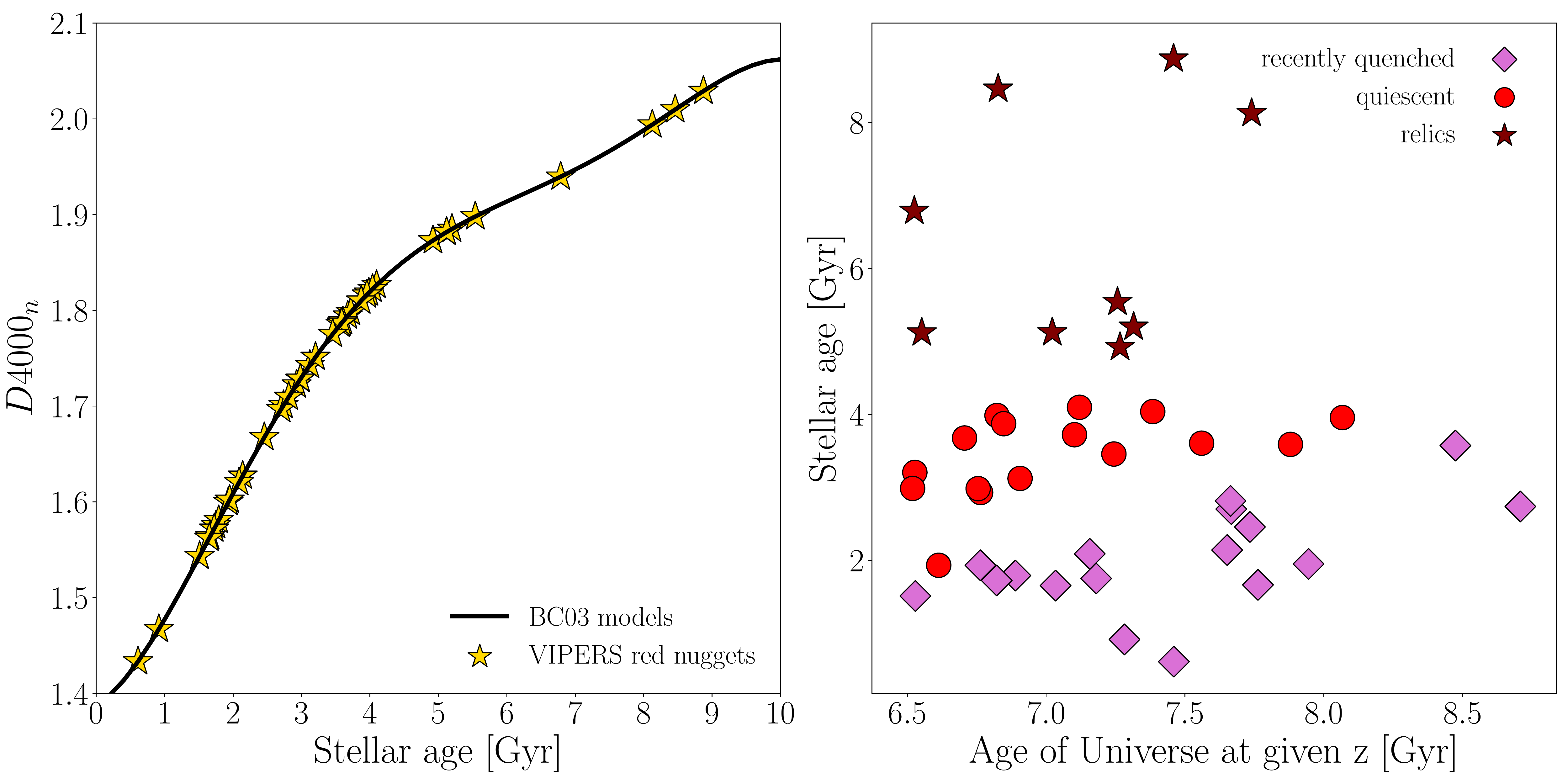}}
	\caption{{Left panel: Stellar age-D4000 relation derived from a grid of synthetic spectra (BC03 model). The model assumes a Chabrier IMF, a single stellar burst with timescales $\tau=0.2$ Gyr and solar metallicity. The  \textit{red nugget sample} is marked with gold stars. Right panel: Stellar ages related to the age of the Universe at given redshift for the \textit{red nugget} sample.}}
	\label{fig:BC03_ages}
\end{figure*}

{Stellar age is one of the most important galaxy physical parameters, but at the same time one of the most difficult to estimate correctly~\citep[e.g.][]{siudek2017, Thomas2019}. There are a number of definitions and methodologies for how to derive stellar ages. In this work, we use the same approach as in~\cite{siudek2017} based on the synthetic D4000-stellar age relation. The synthetic spectra from \cite{Bruzal2003}, hereafter BC03, models are degraded to the typical VIPERS spectral resolution of~14\AA. The grid of synthetic spectra was generated under the assumption of the \cite{Chabrier2003} IMF and single burst SFH with a timescale of the $\tau=0.1, 0.2, 0.3$ Gyr for stellar ages in the range from 1 to 10 Gyr, with steps of 0.25 Gyr. In~\citealt{siudek2017} we obtained a nominal D4000-stellar age relation for BC03 models with different metallicities, $\rm{log(Z/Z_{\odot})=0.4, 0.0, -0.4}$. 
High-mass passive galaxies are supposed to be characterised by slightly super solar metallicity on the order of 0.07 and 0.1 $\rm{log(Z/Z_{\odot})}$ for the higher mass bins of $\sim10^{11.1}$, and $\sim10^{11.4}$ $\rm{M_{star}}$, respectively~\citep{Gallazzi2014, siudek2017}.  However, we are not able to clearly distinguish the effects of stellar ages and this would result in the expected change of D4000 of $\pm0.06$~\citep{siudek2017}, which is less than the average error of the D4000 measurements ($0.08$) for the \textit{red nugget sample}. 
Moreover, properties to select our red nugget sample ($\rm{M_{star}}$, SFRs) were also derived assuming solar metallicity~(see Sec.~\ref{sec:physical_properties}).  
In this paper, we also assume solar metallicity to estimate stellar ages for our sample. The stellar ages obtained from a synthetic D4000-stellar age relation are shown in Fig.~\ref{fig:BC03_ages} and given in Tab.~\ref{table:stellar_ages} in App.~\ref{app:stage_validation}.
}

{As VIPERS galaxies are observed at the intermediate redshift when the Universe was at half of its age, we cannot expect to find stellar ages of $\sim10$ Gyr, but rather ages adequate to the age of the galaxy at the given redshift ($\rm{t_z}$). As relics are believed to be survivors of the first galaxies, we assume that our galaxies formed within the first 0.5 Gyr after the Big Bang~\citep{Naidu2022}. The SFH of red galaxies is modelled with an exponential rise in their SFR of the short timescale~\citep[$\rm{\tau\sim0.2}$ Gyr; e.g.][]{siudek2017}, followed by the quenching time, $\rm{t_Q}$ when galaxy transform from active to passive phase. However, the quenching timescale may vary (broadly assuming from $\sim2$ to $\sim4$ Gyr) depending on the properties of the galaxies themselves, as well as the environment in which they reside~\citep[e.g][]{Wright2019, Bravo2023}. Stellar populations of galaxies that avoided merger processes should be then characterised by the ages (roughly) similar to the age of the galaxy at a given redshift, i.e $\rm{age_{stellar}\sim t_z-(t_Q+\tau})$\footnote{The incorporation of $\rm{\tau}$ or not in this equation does not influence the definition of our samples.}.  
We define \textit{relics} as galaxies with $\rm{age_{stellar}\gtrsim t_z-t_Q}$ assuming rapid quenching  ($\rm{t_Q=2}$ Gyr). 
We further distinguish \textit{recently quenched galaxies} with $\rm{age_{stellar}\lesssim t_z-t_Q}$ assuming $\rm{t_Q=4}$ Gyr and in-between \textit{quiescent galaxies}. For our \textit{red nugget sample} this translates into the stellar ages of $\gtrsim 5 $ Gyr, $\sim3-5$ Gyr, and $\lesssim 3 $ Gyr, for \textit{relics}, \textit{quiescent} and \textit{recently quenched galaxies}, respectively (see the right panel in Fig.~\ref{fig:BC03_ages}). Among 42 \textit{red nuggets} we find 17 \textit{recently quenched}, 16 \textit{quiescent}  and 9 \textit{relic galaxies}. 
We note, that this division should be used with caution, as stellar ages can be affected by a number of factors (e.g. metallicity, SFH, environment) while estimates based on D4000 are uncertain and should be rather combined with other age indicators, such as $H\delta$. 
However, due to the resolution of VIPERS spectra, we cannot measure $H\delta$~(\citealt{siudek2017}). 
Furthermore, we cannot accurately estimate the age of galaxies, which undermines the reliability of \textit{recently quenched} and \textit{quiescent galaxies}. The conclusion should be rather drawn from 9 strong \textit{relic} candidates that are survivors of the first galaxies that avoided the merger process at least until redshift at which were observed. 
In Sec.~\ref{sec:relics_enironmnet}, we investigate the environments of these strong \textit{relic} candidates with respect to the \textit{quiescent} and \textit{recently quenched galaxies}.  
We also note that all of our \textit{red nuggets} are red and massive with no signs of star formation activity, suggesting that they host rather old stellar populations~\citep{Kauffmann2003, Gallazzi2006}.  
For example,~\cite{Wu2020} found that recently quenched systems at $z\sim0.8$ are compact but preferably \textit{bluer} in colour, which differs from the colours of our \textit{red nugget sample}. Due to uncertainties of the stellar age estimates  throughout the main paper, we investigate the environmental properties of \textit{red nuggets}, rather than describing the environments depending on their evolutionary state. }

\subsection{Control samples}\label{sec:control_sample}

To characterise the environment in which \textit{red nuggets} reside we compare them to other red galaxies. 
We limited our analysis to red galaxies only, creating four subsamples based on their stellar masses and radii, as stellar mass enters in defining the environment where these galaxies reside~\citep[e.g.][]{Siudek2022}.  
We generated four control samples: (1) \textit{red sample}, representing the whole variety of the red galaxy population, (2) \textit{red normal-size sample}, limiting the red galaxies to massive, passive and \textit{normal-size} galaxies, i.e normal-size \textit{red nugget} counterparts, (3) \textit{red compact sample}, selecting compact, massive red galaxies, i.e compact-size \textit{red nugget} counterparts, (4) \textit{red intermediate-mass sample}, restricting the red galaxies to \textit{lower masses}, passive and ultra-compact galaxies, i.e intermediate-mass (with $\rm{10.4\lesssim log(M_{star}/M_{\odot})<10.9}$) \textit{red nugget} counterparts. 

The control samples are drawn from the \textit{VIPERS sample} as defined in Sec.~\ref{sec:sample_selection} after removing \textit{red nuggets}, i.e. from the mass-complete sample of 7\,264  galaxies with secure redshifts, sizes and environmental measurements. 
In the following sub-sections, we give a short description of each control sample.

\subsubsection{Red sample}\label{sec:red_sample}

The \textit{red sample} consists of 4\,558 red galaxies selected from \textit{VIPERS sample} based on the NUVrK diagram~\citep{moutard2016b}.  
See App.~\ref{app:colours} for a more detailed description of the colour selection. 
 
Properties of the \textit{red sample} represent the entire population of red galaxies, galaxies that have just joined the passive path as well as the reddest and oldest ones. 
Properties of VIPERS red galaxies  
have already been extensively discussed~\citep{Gargiulo2017,siudek2017,Siudek2018b,Siudek2018a} including determining their environmental dependence \citep{Gargiulo2019, Siudek2022}. 
{Distributions of the main properties ($\rm{M_{star}}$, $\rm{R_e}$ and redshift) of the \textit{red sample} are shown in the first row in Fig.~\ref{fig:sample_param_dif}.}

\subsubsection{Red normal-size sample}\label{sec:redlarge_sample}

We generated a sample of the red massive normal-size galaxies (normal-sized red nugget counterparts), hereafter the \textit{red normal-size sample}. 
The \textit{red normal-size sample} is characterised by similar properties (\textit{$\rm{M_{star}}$}, colours, redshift and SFRs) as the \textit{red nugget sample} but for normal-size galaxies ($\rm{R_e> 2.5}$ kpc; {see the second row in Fig.~\ref{fig:sample_param_dif}}). 
Due to different definitions of compactness ranging from 1.5 to 2.5 kpc, we kept a lower limit of normal-size red massive galaxies above this range. 
The \textit{red normal-size sample} was drawn from the \textit{VIPERS sample} by selecting 875 galaxies covering the same $\rm{M_{star}}$ ($\rm{10.9 \leq log(M_{star}/M_{\odot}) \leq 11.1}$), redshift, colours (NUV$rK$ selection), 
and SFR range ($\rm{SFR<1}$ $\rm{M_{\odot}yr^{-1}}$) as the \textit{red nuggets}, but with larger sizes given by their radii, $\rm{R_e>2.5}$ kpc. 
The ranges of these parameters were not strictly defined during the  \textit{red nugget} selection (see Sec.~\ref{sec:rednugget_sample}), but roughly correspond to the ranges found for \textit{red nugget sample} {(see Fig.~\ref{fig:rednugget_validatio})}. 
To even more accurately recreate the $\rm{M_{star}}$ distribution of the \textit{red nugget sample} {(see the second row in the first column in Fig.~\ref{fig:sample_param_dif})}, we select only galaxies with the closest $\rm{M_{star}}$ estimate among the sample of 875 large ($\rm{R_e>2.5}$ kpc) red massive galaxy population. 
Our final \textit{red normal-size sample} consists of 42 galaxies that recreate the stellar mass distribution of \textit{red nugget sample}. 

\subsubsection{Red compact-size sample}\label{sec:redcompact_sample}

As mentioned in the previous Section, the upper limit for selecting compact galaxies varies from 1.5 to 2.5 kpc. Thus, we also created a sample of red massive galaxies with a less restrictive cut on the size than the one applied for \textit{red nugget sample}. We drew \textit{red compact-size sample} from the \textit{VIPERS sample} by selecting 266 massive ( $\rm{log(M_{star}/M_{\odot}) \geq 10.9}$) galaxies covering the same redshift, colours ($\rm{r-K}$ and $\rm{NUV-r}$), 
and SFR range ($\rm{log(SFR)<0}$ $\rm{M_{\odot}yr^{-1}}$) as \textit{red nuggets} but with intermediate sizes given by their radii, $\rm{1.5<R_e<2.5}$ kpc {(see the third row in Fig.~\ref{fig:sample_param_dif})}. This sample represents \textit{red nuggets} defined by more relaxed criteria on their sizes and corresponds rather to the definitions proposed by~\citealt{Barro2013} and~\citealt{Damjanov2015} at the high-mass end. At the same time, this sample represents galaxies with high surface mean stellar mass density, $\rm{\Sigma=M_{star}/(2\pi R_{e}^2)>2000}$ $\rm{M_{\odot}pc^{-2}}$. In this paper, a sample of high-$\Sigma$ red galaxies is used only for comparison of environments of \textit{red nuggets}. The properties of low- and high-$\Sigma$ massive passive galaxies have already been studied by~\citealt{Gargiulo2017} and their environmental properties have been extensively investigated by~\citealt{Gargiulo2019}. The comparison of the \textit{red nugget sample} with the \textit{red compact-size sample} may show whether a more restrictive definition of compactness ($\rm{R_{e}<1.5}$ kpc) changes the environmental properties of red massive compact galaxies. 

\subsubsection{Red intermediate-mass sample}\label{sec:redlowmass_sample}

\begin{figure*}
 	\centerline{\includegraphics[width=0.99\textwidth]{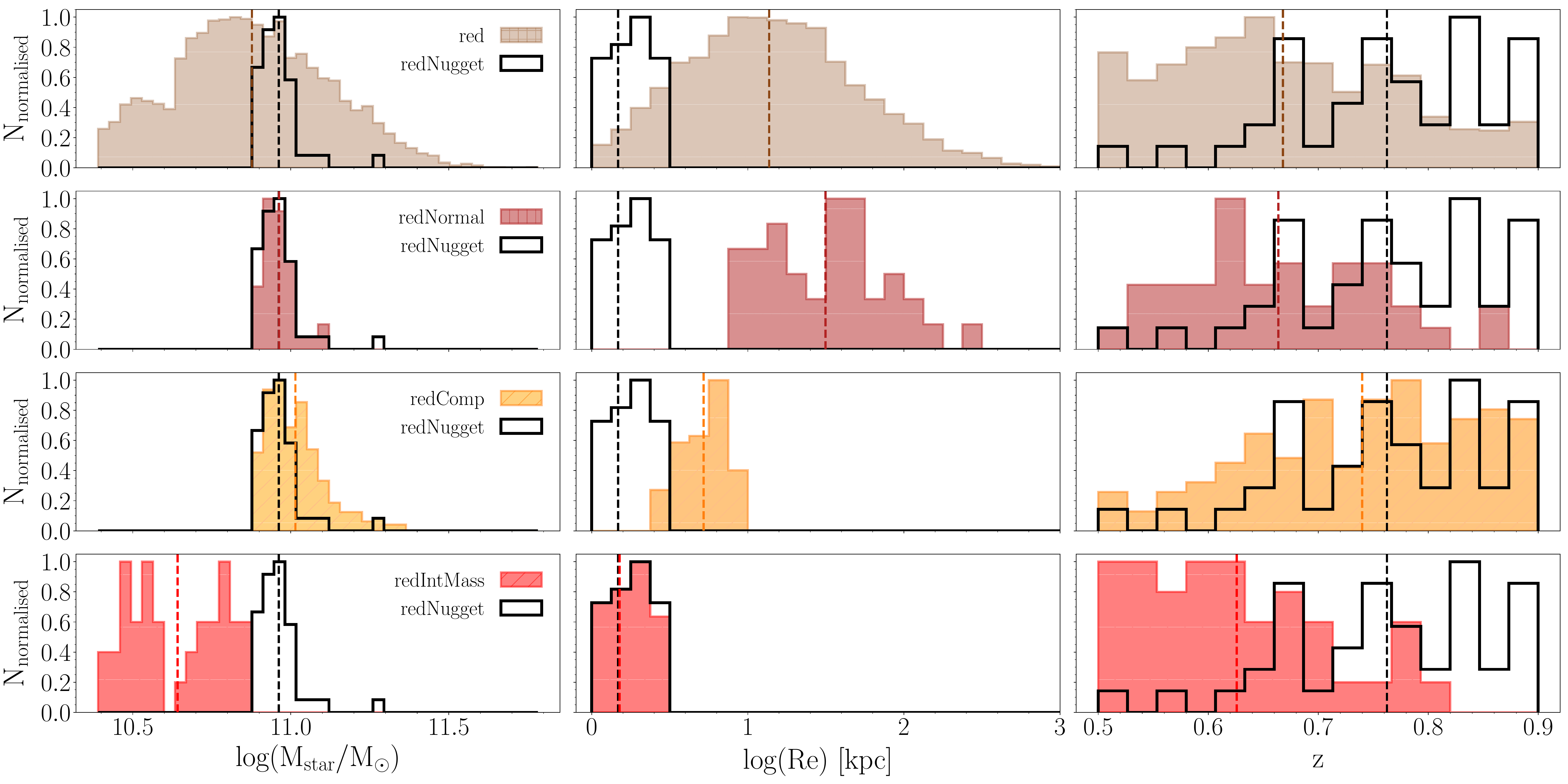}}
	\caption{Distributions of $\rm{M_{star}}$ (left), $\rm{R_e}$ (centre), and redshift (right) of the  \textit{red nugget sample} (in black),  \textit{red sample} (in brown), \textit{red compact-size sample (in orange)}, \textit{red normal-size sample} (in firebrick) and \textit{red intermediate-mass sample} (in red). The mean values are marked with dashed lines. }
	\label{fig:sample_param_dif}
\end{figure*}

To have a complete picture of the dependence of red galaxies on the environment, we lastly created the sample of intermediate-mass red nugget counterparts, hereafter a \textit{red intermediate-mass sample}. 
The \textit{red intermediate-mass sample} has similar properties (\textit{\rm{$R_e$}}, colours, redshift and SFRs) as the \textit{red nugget sample} but for lower-mass galaxies. 
This sample was drawn from the \textit{VIPERS sample} and consists of 527 galaxies covering the same range of radii, colours ($\rm{r-K}$ and $\rm{NUV-r}$) and redshift as the \textit{red nugget sample}, but it is placed in the lower-mass regime (10.4 $\lesssim\rm{log(M_{star}/M_{\odot})<10.9}$; {see the last row in Fig.~\ref{fig:sample_param_dif}}). 
We further restrict our sample to 42 galaxies with the closest sizes to the sizes of \textit{red nuggets} to recreate the $\rm{R_e}$ distribution of the \textit{red nugget sample}. 

Properties of the samples, namely their average $\rm{M_{star}}$, $\rm{R_e}$ and SFRs are given in Tab.~\ref{table:control_sample_properties}.  
The comparison of the distributions of their selection-based properties is shown in Fig.~\ref{fig:sample_param_dif} and the p-values of the KS test are given in the Tab.~\ref{table:Kstest}.  
The control samples are similar to \textit{red nugget sample} in SFR, r-K and NUV-r colours, as all of these samples represent the red passive population. However, 
the \textit{red normal-size sample} is the only sample recreating the $\rm{M_{star}}$ distribution of the \textit{red nugget sample}, while only the distribution of sizes of the \textit{red intermediate-mass sample} fits the ones of \textit{red nuggets}. 
At the same time, the \textit{red sample} covers a wider range of $\rm{M_{star}}$ and $\rm{R_e}$ representing the diversity of red galaxies and does not recreate the distributions of $\rm{M_{star}}$ and $\rm{R_e}$ of the \textit{red nugget sample}. 
Similarly, the \textit{red compact-size} sample does not mirror the $\rm{M_{star}}$ and $\rm{R_e}$ distributions of the \textit{red nugget sample}. 
The colours of the \textit{red nugget} and control samples are evaluated in App.~\ref{app:colours}. 

\begin{figure}
	\centerline{\includegraphics[width=0.49\textwidth]{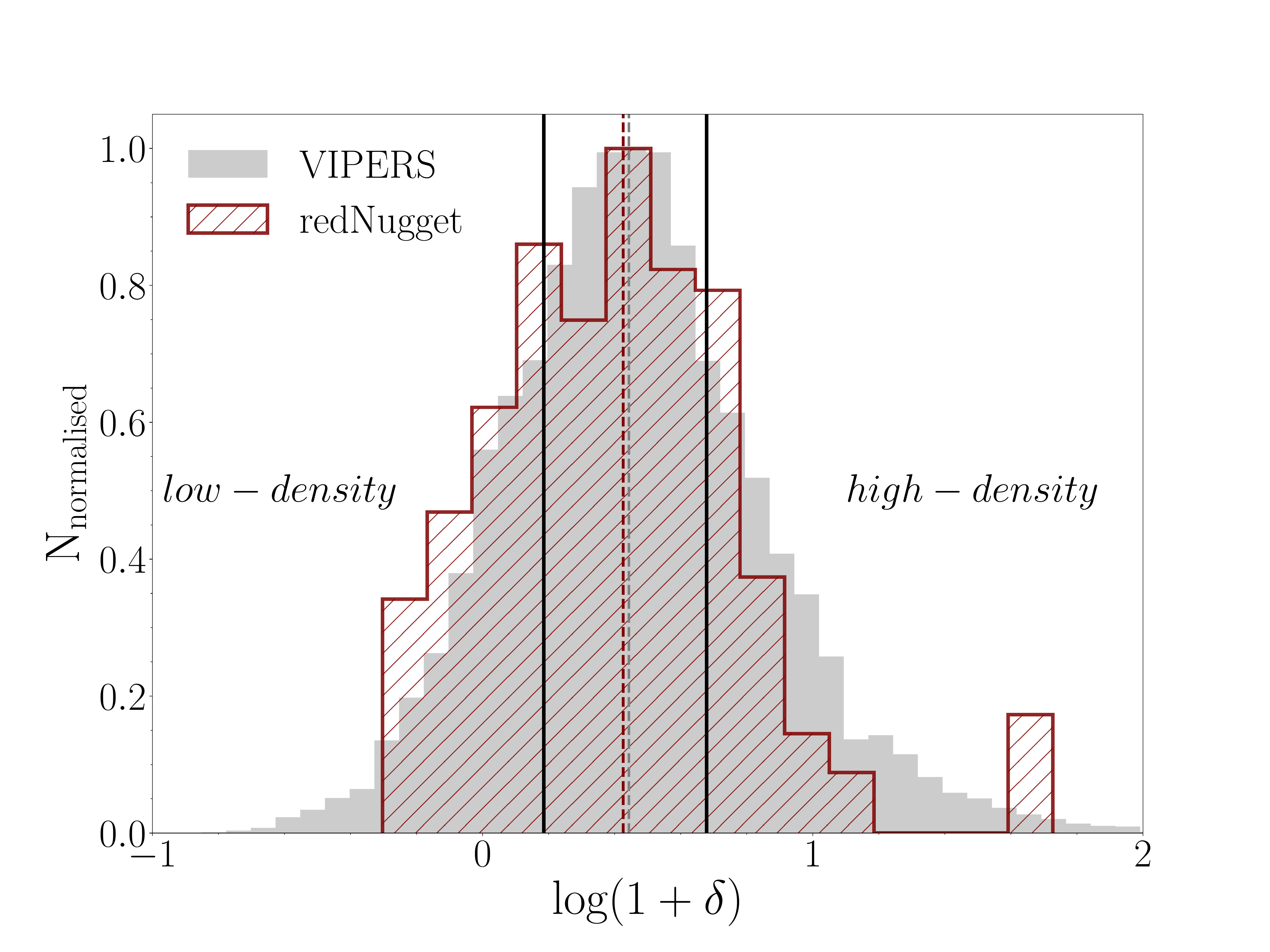}}
	\caption{Density distributions for the \textit{VIPERS sample} and \textit{red nuggets}. The first and third quartile of the VIPERS $\delta$ distribution used to define the LD and HD environments are marked with solid black lines. The dashed lines correspond to the mean values.}
	\label{fig:density_histo}
\end{figure}

	\begin{table}
		\centering                         
		\begin{tabular}{P{1.5cm} P{0.3cm} P{1.0cm} P{1.0cm} P{1.0cm} P{1.3cm} }    
			\hline 
			sample & N  & z & $\rm{M_{star}}$ & $\rm{R_{e}}$ & SFR \\
			\hline 
            \hline
			\textit{redNugget} & 42 & 0.67$\pm$0.10 & 10.96$\pm$0.06 & 1.21$\pm$0.25 & -4.84$\pm$2.74   \\
			\textit{red} & 4\,558 & 0.67$\pm$0.10 & 10.88$\pm$0.24 & 3.72$\pm$2.71 & -4.04$\pm$2.75   \\
			\textit{redNormal} & 42 & 0.66$\pm$0.09 & 10.96$\pm$0.05 & 4.81$\pm$1.94 & -3.51$\pm$2.71   \\
			\textit{redCompact} & 266 & 0.74$\pm$0.10 & 11.01$\pm$0.09 & 2.07$\pm$0.28 & -5.00$\pm$2.61   \\
			\textit{redIntMass} & 42 & 0.63$\pm$0.08 & 10.64$\pm$0.15 & 1.22$\pm$0.23 & -5.24$\pm$2.48   \\
			\hline
		\end{tabular}
		\caption{Physical properties of the \textit{red nugget sample} and control samples: \textit{red sample}, \textit{red normal-size sample}, \textit{red compact-size sample}, \textit{red intermediate-mass sample}. The number of galaxies (N) and the mean with standard deviation for: redshift, z, stellar masses, $\rm{M_{star}=log(M_{star}/M_{\odot})}$, radii, $\rm{R_{e}}$ [kpc], and $\rm{SFR=log(SFR)}$ $\rm{[M_{\odot}yr^{-1}]}$ are provided. 			 }             
		\label{table:control_sample_properties}     
	\end{table}

	\begin{table}
		\centering                         
		\begin{tabular}{P{1.0cm} P{0.85cm} P{0.85cm} P{0.85cm} P{0.7cm} P{0.85cm} P{0.70cm} }    
			\hline 
			sample & z & $\rm{M_{star}}$ & $\rm{R_{e}}$ & SFR & r-K & NUV-r  \\
			\hline 
            \hline
			\textit{red} & $\leq0.01$ & $\leq0.01$ & $\leq0.01$ & \textbf{0.23} & \textbf{0.53}  & \textbf{0.19}   \\
			\textit{redNormal} & $\leq0.01$ & \textbf{0.99} & $\leq0.01$ & \textbf{0.09} & \textbf{0.39} & \textbf{0.26} \\
			\textit{redCompact} & \textbf{0.59} & $\leq0.01$ & $\leq0.01$ & \textbf{1.00} & \textbf{0.33} & \textbf{0.49}   \\
			\textit{redIntMass} & $\leq0.01$ & $\leq0.01$ & \textbf{1.00} & \textbf{0.91} & \textbf{0.56} & \textbf{0.99}  \\
			\hline
		\end{tabular}
		\caption{P-values of the KS test of differences between the distributions of properties of the \textit{red nugget} and \textit{control samples}: \textit{red sample}, \textit{red normal-size sample}, \textit{red compact-size sample} and \textit{red intermediate-mass sample}. The KS test returns the probabilities under the null hypothesis that the two samples are drawn from the same population. The p-values indicating that the null hypothesis is rejected (i.e distributions are \textit{not} drawn from the same population) have low values ($\rm{p_{KS}<0.05}$). The p-values above the significance level ($\rm{p_{KS}>0.05}$) indicate that there are no statistically significant differences in the distributions of the two samples. The probabilities with a significant level above our threshold are shown in bold.	 }             
		\label{table:Kstest}     
	\end{table}
	
\begin{figure*}
  	\centerline{\includegraphics[width=0.99\textwidth]{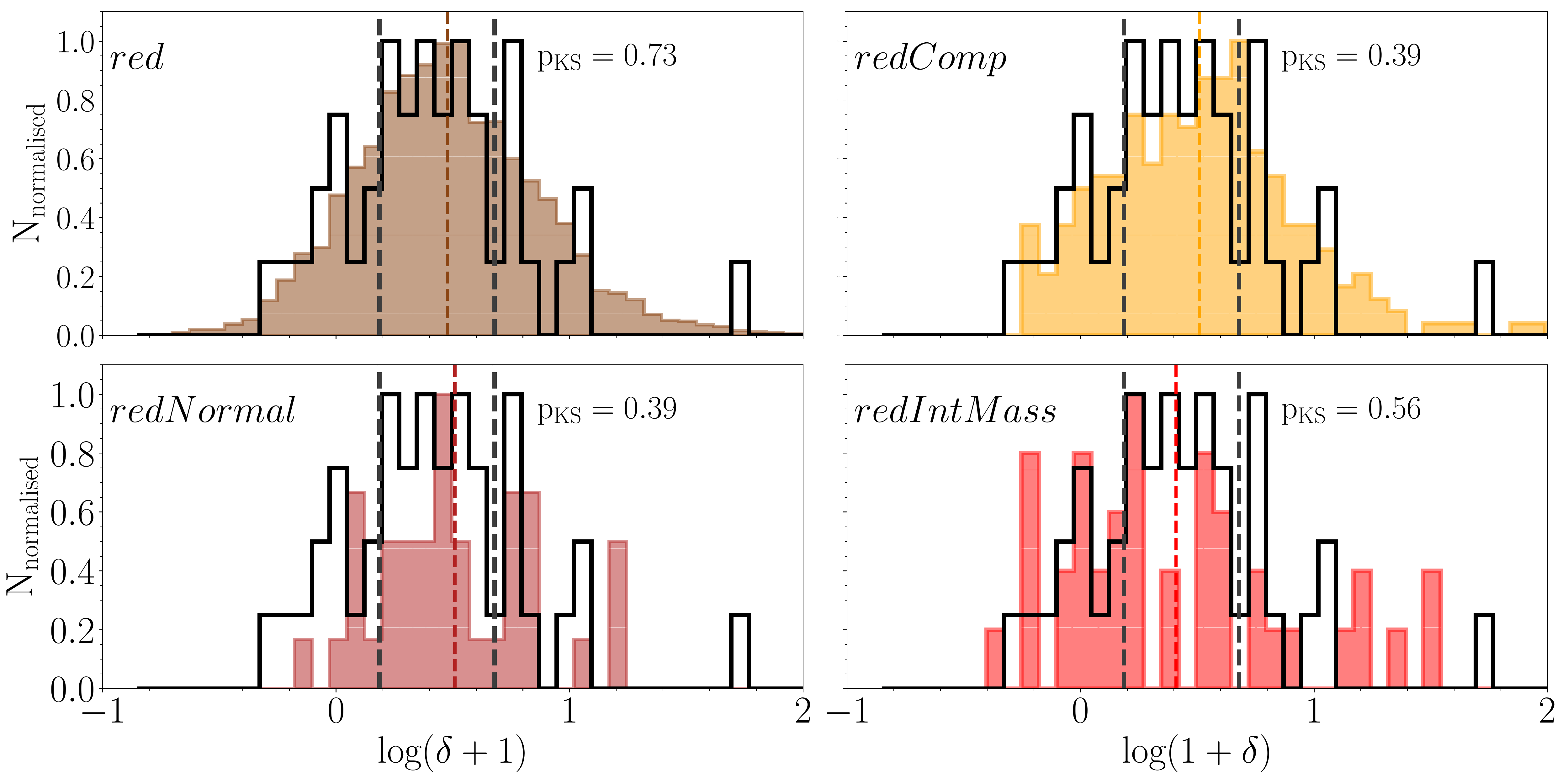}}
	\caption{Density distributions for the control galaxy samples and the \textit{red nugget sample} (in black). The first and third {quartiles} of the VIPERS $\delta$ distribution (corresponding to LD and HD, respectively) are marked with dashed grey lines. The dashed lines correspond to the mean values, and probabilities of the KS test are reported in the plots. }
	\label{fig:density_control_histo}
\end{figure*} 

\section{Environment of red nuggets}\label{sec:environment_properties}

In the following Sections we investigate how the fractions of \textit{red nuggets} and red control galaxies vary as a function of the environment, as quantified by the density contrast, $\rm{\delta}$ (see Section~\ref{sec:environment}).  
 
\subsection{Density distributions}\label{sec:density_distribution}

The $\rm{log(1+\delta)}$ distribution for the \textit{VIPERS sample} (i.e. including both blue galaxies preferably found in LD environments as well as red galaxies found mostly in HD environments) is shown in Fig.~\ref{fig:density_histo}. 
The thresholds for LD and HD environments are defined as the first and third quartiles of the $\rm{log(1+\delta)}$ distribution of these galaxies and have values of 0.2 and 0.7, respectively. 
These thresholds are in agreement with those found by~\citealp{Davidzon2016} and~\citealp{Cucciati2017}, with different sample sizes (the first VIPERS public release but not accounting for reliable size measurements) and different SED fitting techniques. 
This confirms the consistency of the VIPERS $\delta$ measurements, independent of the sample cuts used in different analyses. 
Those thresholds may be compared to the large-scale structure where LD environments would correspond to void galaxies (with the average projected $\rm{D_{p,5}}$ $\sim3.5\rm{h^{-1}Mpc}$), and HD would reach (although the dimension is still too high) group and cluster members (with the average projected $\rm{D_{p,5}}$ $\sim2\rm{h^{-1}Mpc}$;~\citealp{Cucciati2017}). 


Among 42 \textit{red nuggets} we found 10 in HD environments and 11 in LD environments. 
The uniform distribution of \textit{red nuggets} over a range of different environments suggests that their frequency is independent of environment. 
There is only one \textit{red nugget} galaxy residing in the high-$\delta$ tail of VIPERS distribution (with $\rm{log(1+\delta)=1.72}$). However, this galaxy does not have any property that is outstanding from the average properties found for the \textit{red nugget sample}. In particular, the galaxy is a high-mass ($\rm{log(M_{star}/M_{\odot})=10.95}$), compact ($\rm{R_e=1.3}$ kpc) object at redshift $\rm{z=0.75}$ with old stellar ages of $\sim5$ Gyr (see the second row in Tab.~\ref{table:stellar_ages} for $\rm{VIPERS=103137838}$).

We checked whether the environmental dependence of \textit{red nuggets} is different than those observed for different representative samples of red galaxies. 
The $\rm{log(1+\delta)}$ distributions for the \textit{red nugget sample} and control galaxy samples are shown in Fig.~\ref{fig:density_control_histo}.  
The $\rm{log(1+\delta})$ distribution of \textit{red nuggets} is similar to all control samples as indicated by 
the KS test (with $\rm{p_{KS}}\gtrapprox0.4$ we cannot reject the hypothesis that the distributions are drawn from the same sample). 

\subsection{The fraction-density relation}\label{sec:fraction_environment_relation}

\begin{figure}
	\centerline{\includegraphics[width=0.49\textwidth]{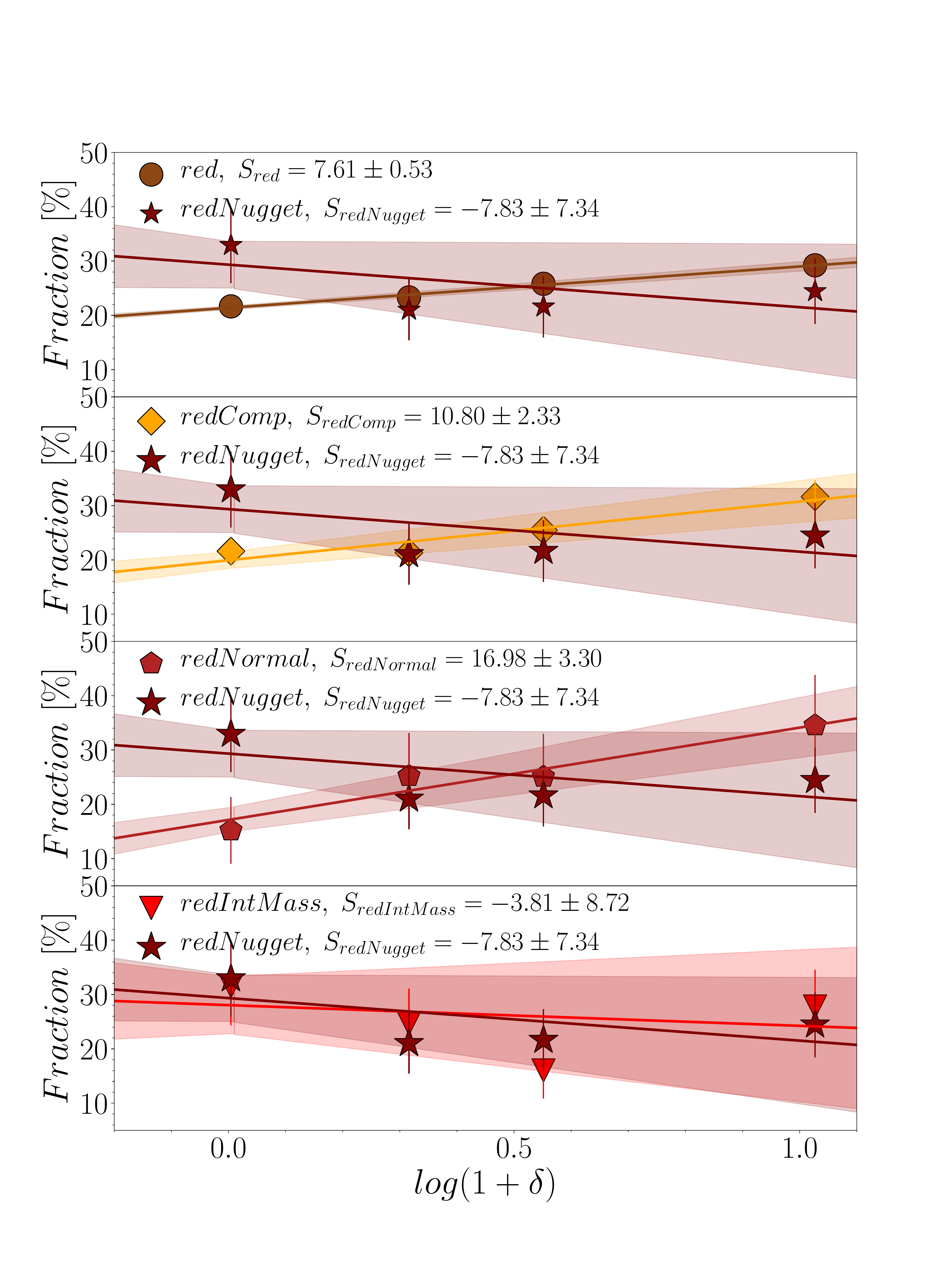}}	
	\caption{Fractions of the \textit{red nugget sample} and control samples as a function of $\rm{log(1+\delta)}$. The centres of $\delta$-bins correspond to the means of the quartiles of the $\delta$ distribution for the \textit{VIPERS sample}. The fractions are normalised to the total number of galaxies in a given sample, i.e. for each sample the fractions in $\delta$ bins sum up to 100\%. The solid line corresponds to the weighted fit. Shaded stripes around lines display the $1\sigma$ uncertainties of the fit. The error bars correspond to the Poisson error. The slope of the fit is given in the legend. }
	\label{fig:fraction}
\end{figure}

To study the environmental trends for \textit{red nuggets}, the relation between the fraction of \textit{red nuggets} and the environment is shown in Fig.~\ref{fig:fraction}. 
For clarity, the centres of the $\delta$ bins correspond to the means of the quartiles of the $\delta$ distribution of the \textit{VIPERS sample}. 
The fraction for each sample is normalised to the total number of galaxies in a given sample, i.e. the fraction sums up to 100\% for each sample. 
Such a normalisation allows us to directly compare the slopes of the fraction-$\delta$ relation between different samples. 
The correlation of the \textit{red nuggets} and \textit{control samples} with the environment is also expressed as the change of their number in the LD and HD environments reported in Tab.~\ref{table:number_in_HD_LD}. 
For the \textit{red sample} the number of galaxies increases by $26\pm11\%$ when moving from LD to HD environments. 
The overabundance is even more pronounced for the \textit{red compact-size} and \textit{red normal-size samples}, as their number in the HD environment is increasing by $47\pm12\%$ and $86\pm14\%$, respectively, with respect to the ones residing in the LD environments. 
The opposite trend is found for the \textit{red intermediate-mass sample} for which the number of galaxies decreases by $33\pm8\%$ when moving to HD environments. At the same time, the fractional decrease by $9\pm10\%$ in the number of \textit{red nuggets} residing in HD with respect to LD regions is consistent with a flat distribution (see Tab.~\ref{table:number_in_HD_LD}).

	\begin{table}
		\centering                         
		\begin{tabular}{P{2.5cm} P{1.cm} P{1.cm} P{1.cm}}    
			\hline 
			sample  & $\rm{N_{LD}}$ & $\rm{N_{HD}}$ & $\rm{\Delta}$ [\%]\\
			\hline 
            \hline
			\textit{redNugget} & 11 & 10 & $\mathbf{-9\pm10}$  \\
			\textit{red} & 1\,023 & 1\,290 & $\mathbf{26\pm11}$  \\
			\textit{redNormal} & 7 & 13 & $\mathbf{86\pm14}$   \\
			{\textit{redCompact}} & {58} & {85} & $\mathbf{47\pm12}$   \\
			\textit{redIntMass} & {15} & {10} & $\mathbf{-33\pm8}$   \\
			\hline
		\end{tabular}
		\caption{Number of galaxies for \textit{red nuggets} and control samples found in LD and HD, environments and the percentage change between them. 
			 }             
		\label{table:number_in_HD_LD}     
	\end{table}

Environmental trends for \textit{red}{, \textit{red compact-size} and} \textit{red normal-size samples} are different from the ones found for \textit{red nuggets} and \textit{red intermediate-mass samples} as expressed also by the slopes of the fraction-$\delta$ relation shown in Fig.~\ref{fig:fraction}. 
The steepest positive relation is found for the \textit{red normal-size sample} {($\rm{S_{redNormal}=16.98\pm3.30}$)} revealing its strong preference for denser environments. 
The {\textit{red compact-size} and }\textit{red samples} also show positive trends, but the slope is $\sim$two times smaller than for the \textit{red normal-size sample} {($\rm{S_{redCompact}=10.80\pm2.33}$, $\rm{S_{red}=7.61\pm0.53}$)}. 
On the contrary, the \textit{red intermediate-mass sample} and \textit{red nuggets} show negative trends for the fraction-$\delta$ relation, consistent within $1\sigma$ ($\rm{S_{IntMass}=-3.81\pm8.72}$, $\rm{S_{redNugget}=-7.83\pm7.34}$). 
In fact, the gradient in the \textit{red nugget} fraction is consistent within $\rm{1-2\sigma}$ with the trends found for the other samples except \textit{red normal-size} galaxies. 
This suggests that \textit{red nuggets} occupy different environments {on average than} \textit{red normal-size} galaxies. In particular, these findings suggest that the positive relation for the \textit{red sample} is driven by {\textit{red normal-size} and \textit{red compact-size} galaxies}, but this trend is tempered by red {ultra-compact galaxies (\textit{red nugget and red intermediate-mass samples})}. 

We verify that our findings are independent of the tracer used to compute the $\delta$ field. 
Considering brighter tracers, $\rm{M_{B} \le (-20.9 - z)}$, which remain complete up to a redshift $\rm{z =1.0}$, the \textit{red nugget sample} increases by 42\%.  
The \textit{red nuggets} sample built with these brighter tracers consists of 60 objects with reliable environmental measurements.  
At the same time, our findings are preserved: \textit{red nuggets} occupy both LD and HD environments, recreating the same fraction-$\delta$ relation as shown in Fig.~\ref{fig:fraction}. 
The only difference with reference to our main \textit{red nugget sample} is a change in the slope of $<0.1\sigma$ ($\rm{S_{redNugget}=-6.91\pm11.25}$). 
Despite better statistics, throughout the paper, we are following the local environment calculated using a fainter tracer, $\rm{M_{B}\leq -20.4-z}$. 
This cut narrows the redshift range for our analysis (and our sample), but rejects possible discreteness evolutionary effects, which can bias our results~\citep{Cucciati2017}. Nevertheless, the performed test can be treated as a sanity check, as both tracers give coherent conclusions. 
{We also verify that changing the binning of the $\delta$ distribution (taking as a reference the $\delta$ distribution of the \textit{red sample} or using deciles of the $\delta$ distribution) does not alter our conclusions. }

 \subsection{Environment of \textit{relic candidates}}\label{sec:relics_enironmnet}
 
 {As shown in Sec.~\ref{sec:rednugget_ages} and App.~\ref{app:stage_validation}, only $\sim20\%$ of our \textit{red nugget sample} (9 out of 42) consists of galaxies with stellar ages almost as old as the Universe at given redshift. 
 According to the KS test ($\rm{p_{KS}=0.24}$; see Tab.~\ref{table:HD-LD_number_Age}), we cannot reject the hypothesis that the $\delta$ distributions of the \textit{red nugget sample} and the \textit{relic sample} are drawn from the same parent distribution. 
 On the other hand, we found 5 \textit{relics} in high-$\delta$ regions and only one in low-$\delta$ environments (see Tab.~\ref{table:HD-LD_number_Age}), suggesting that \textit{relics} are preferably found in high-$\delta$ environments. 
 At the same time, \textit{quiescent galaxies} are overabundant in low-$\delta$ environments and \textit{recently quenched} galaxies are distributed both in low- and high-$\delta$ regions. 
 This suggests that the stellar populations of \textit{red nuggets} become older when moving to denser regions. 
 }
	\begin{table}
		\centering                         
		\begin{tabular}{P{2.5cm} P{1.cm} P{1.cm} P{1.cm} P{1.cm}}    
			\hline 
			sample  & $\rm{p_{KS}}$ & $\rm{N_{LD}}$ & $\rm{N_{HD}}$ & $\rm{\Delta}$ [\%]\\
			\hline 
            \hline
			\textit{redNugget} &  & 11 & 10 & $-9\pm10$  \\
			\textit{relics} & 0.24 & 1 & 5 & $400\pm22$  \\
			\textit{quiescent} & 0.83 & 7 & 2 & $-71\pm5$  \\
			\textit{recently quenched} & 0.99 & 3 & 3 & $0\pm10$   \\
			\hline
		\end{tabular}
		\caption{{Number of galaxies in \textit{red nugget sample} and its subsamples distinguished according to their stellar ages: \textit{relic}, \textit{quiescent} and \textit{recently quenched galaxies}, found in LD and HD, environments and the percentage change between them. The KS test of $\delta$ distribution for \textit{red nugget sample} and its subsamples is reported.}  
			 }             
		\label{table:HD-LD_number_Age}     
	\end{table}

\section{Comparison with the literature}\label{sec:comparison_literature}

Counter-intuitively, the first low-redshift observations found red nuggets in clusters. 
\cite{poggianti2013} found that local massive compact galaxies (defined to have densities $\rm{\geq 3 \times 10^9 M_{\odot}kpc^{-2}}$) from the WINGS sample are three times more frequent in clusters with velocity dispersions above 500 $\rm{km s^{-1}}$ than in the field.  
However, using the Millennium I-WMAP7 simulations, \cite{Peralta2016} found that massive ($\rm{log(M_{star}/M_{\odot})>10}$) relics (defined as a galaxy that has increased its stellar mass by less than 10\% since z$\sim$2) can be found in all type of environments.  
Nevertheless, their results suggest that the fraction of massive relics increases up to 4.6 times (from a theoretical point of view) when moving to HD environments. 
In the same work, \citealp{Peralta2016} compared their results obtained from simulations with observed data from the New York University Value-Added Galaxy Catalogue based on the 7th release of the Sloan Digital Sky Survey. 
They found a similar trend, with relics defined as massive and old ETGs ($\rm{log(M_{star}/M_{\odot})>10}$,
age $>$10 Gyr) 
constituting $1.11\pm0.05\%$ among massive galaxies in LD environments, increasing their fraction to $2.4\pm0.4\%$ in HD environments.

Those authors~\citep[][]{poggianti2013,Peralta2016} claim that red nuggets prefer the centres of clusters with high velocity dispersions and a very hot intracluster medium.  
Such conditions effectively prevent galaxies from mass and size growth. 
The survival of relics at the centre of clusters can be explained also under the hierarchical model of galaxy formation, as they are a subpopulation of ETGs which have been formed at high redshift, and thus had more time to cluster than galaxies formed later~\citep[][]{Peralta2016}. 
Several other works have confirmed that compact passive galaxies are more likely to be found in HD environments both from a theoretical point of view and in observations~\citep[e.g.][]{Valentinuzzi2010a,Stringer2015,Baldry2021}. 
On the contrary, in this work, we found a negative fraction-$\delta$ relation for red nuggets (constant within $1\sigma$), suggesting that \textit{red nuggets} at least are found in different types of environments or even prefer LD environments. 
However, we note that VIPERS environmental measurements do not catch the central regions of clusters, as the $\rm{D_{p,5}}$ in the highest $\delta$ bin is still too large to be comparable with the small dimensions of galaxy groups and clusters~\citep{Cucciati2017}. 

The VIPERS environmental trend is in agreement with the findings of~\cite{Tortora2020}, who found that the fraction (weighted to the total number of galaxies, i.e. including both normal-size galaxies and ultra-compact massive galaxies) of KIDS red nuggets (defined as galaxies with $\rm{R_{e} < 1.5}$ kpc and stellar masses $\rm{log(M_{star}/M_{\odot})>10.9}$ at $\rm{z<0.5}$)  in clusters is lower than in the field, and it is consistent within the typical uncertainties. 
Moreover, they found that the fraction of ultra-compact massive galaxies in peripheral regions of clusters is doubled with respect to the ones in the central regions. 
Interestingly, \cite{Tortora2020} also showed that the fraction of red nuggets in LD environments has dropped significantly since $\rm{z\sim0.45}$, with only a minor change in the fraction of the most massive ($\rm{log(M_{star}/M_{\odot})>11.2}$) red nuggets in clusters. 
\cite{Tortora2020} found a decrease in the fraction of red nuggets in LD environments, from $\sim1.6\%$ at $\rm{z\sim0.44}$ to $\sim0.3\%$ at $\rm{z\sim0.26}$. 
At the same time, the fraction of red nuggets in HD environments decreases from $\sim0.9\%$ to $\sim0.2\%$ at the same redshift intervals. 
The fraction of VIPERS \textit{red nuggets} normalised to the total number of red  galaxies (to mimic selection used by~\citealp{Tortora2020}) decreases from $1.5\pm0.3\%$ to $0.8\pm0.2\%$ when moving to denser environments. 
Taking into account the increasing trend of the fraction-redshift relation found by~\citealp{Tortora2020} (see the right panel in Fig.~1) and the possible sample selection bias, our estimations support the decreasing fraction of red nuggets with denser environments and are in agreement with trends presented by~\cite{Tortora2020}.  
Similarly,~\citealp{Buitrago2018} found that massive compact galaxies (defined as galaxies with $\rm{R_e<2}$ kpc and $\rm{log(M_{star}/M_{\odot})>10.9}$ at redshift $\rm{0.02 < z < 0.3}$) drawn from the GAMA survey reside in both HD and LD environments. 
\cite{schnorr22} also found that ultra-compact, massive quiescent  galaxies at $\rm{z<0.15}$ drawn from the MANGA survey reside in lower mass halos than their non-compact counterparts.

\begin{figure}
	\centerline{\includegraphics[width=0.49\textwidth]{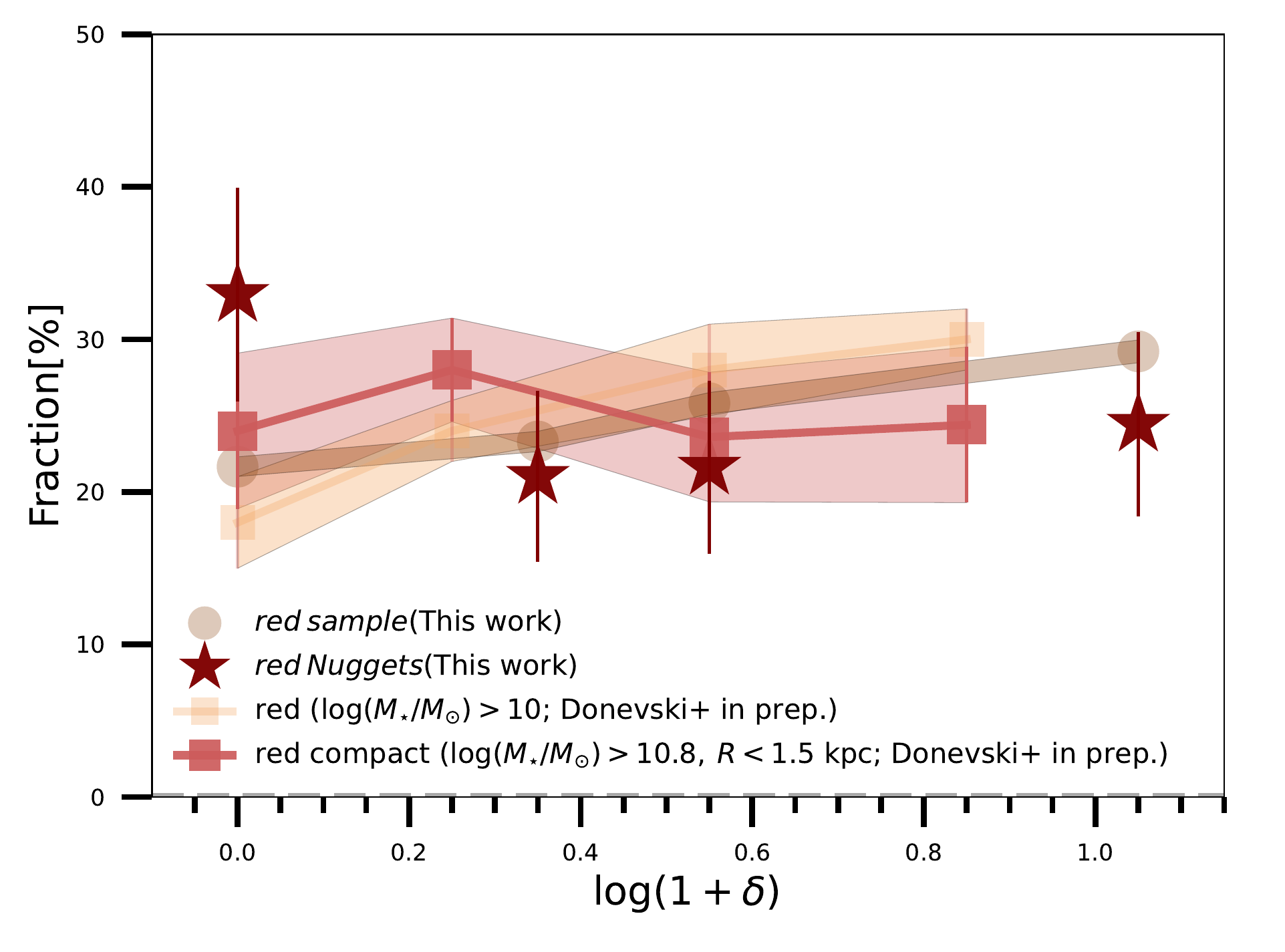}}	
	\caption{The fraction-$\delta$ relation found for the  VIPERS (this work) and COSMOS (Donevski et al. (in prep.)) samples.  }
	\label{fig:donevsky}
\end{figure}

Recently, Donevski et al. (in prep.) studied the connection between local overdensity and the SFHs of 1\,200 quiescent galaxies observed at $\rm{0.1<z<0.7}$ in the COSMOS field. 
Their selection was based on the spectral age index ($\rm{D4000>1.5}$) and specific SFR ($\log(\rm sSFR/yr^{-1})<-10.5)$, while they adopted a weighted adaptive kernel smoothing estimator with a global width of 0.5 Mpc as a probe of the local overdensity field. 
For the entire  quiescent population, they found a positive correlation of the fraction-overdensity relation with a similar trend as we find for the \textit{red sample} (the trends agree within $1\sigma$). 
Similarly to this work, they obtained a mild anti-correlation for the sub-sample of the most compact galaxies ($\rm{\log(M_{star}/M_{\odot})>10.8, R_{e}<1.5}$ kpc). 
Namely, the fraction of red nuggets in Donevski et al. (in prep.) {is flat within $1\sigma$ (declines }from $28\pm4\%$ to $24\pm6\%$ while moving from LD to HD environments; see Fig.~\ref{fig:donevsky}{)}. 
These values agree within $1\sigma$ with our estimates ($33\pm7\%$, $24\pm6\%$ for LD, and HD environments, respectively). 
{
We note, that fraction of \textit{red nuggets} in the LD environment found by Donevsky et al. (in prep.) corresponds rather to the lower limit, as this fraction increases by a few percent when another prescription of the SFH in the SED fitting procedure is assumed. }
Despite the differences between our study and that of Donevski et al. (in prep.) regarding sample selections, observed redshift ranges and adopted overdensity estimators, the inferred overall trends and estimated fractions are highly consistent, both for the general sample of red quiescent galaxies and for the compact red nuggets. 
{Such consistency in the fraction-$\delta$ relation suggests that the possible bias introduced by sample selection or definition of the environment, does not influence our main conclusions.}

\section{Discussion}\label{sec:discussion}

The positive trend of the fraction-$\delta$ relation for red galaxies has been observed at both low and intermediate redshifts~\citep[e.g.][]{Dressler1980,Balogh1997,Kauffmann2004,Cucciati2017,PaulinoAfonso2019, Siudek2022}. 
However, in~\citealt{Siudek2018b,Siudek2018a, Siudek2022Proceeding},  we have shown that the red population is not entirely homogeneous. 
We have distinguished three subclasses of red galaxies, the properties of which behave differently when moving towards denser environments~\citep{Siudek2022}. 
Discrepancies found between the environmental trends of the three red subclasses suggest that the positive trend of the fraction-$\delta$ relation might be driven not by the whole population of red galaxies, but rather by a specific subclass of red galaxies. 
Results presented in this paper support such a hypothesis, as the increasing fraction of red galaxies when moving to denser environments is driven by normal-size {or compact size (with $\rm{1.5<R_e<2.5}$ kpc)} galaxies (the \textit{red normal-size} {and \textit{red compact-size}} sample{s}; see {the second and third panel in }Fig.~\ref{fig:fraction}). 
At the same time, ultra-compact {(with $\rm{R_e<1.5}$ kpc)} galaxies (\textit{red intermediate-mass} and \textit{red nugget samples}) show mild negative trends (see {the last panel in }Fig.~\ref{fig:fraction}), suggesting that the HD environment penalises the {survival} of ultra-compact galaxies independently of their mass. 
Such behaviour seems intuitive as merger events are more frequent in dense environments and ultra-compact galaxies are more likely to be cannibalised~\citep[e.g.][]{Wellons2016}. 
This possibility is also predicted by the analytical model of \cite{lapi18} which proposes that the large scatter in size-stellar mass relation of quiescent galaxies is a consequence of preserving ultra-compact sizes by a fraction of ETGs due to the lack of late-time dry merger events. 
Moreover, these galaxies are expected to host extremely massive black holes at their centres and may have originated a strong puffing-up episode at the peak time of their activity. 
On top of this, \cite{zanisi21} adopted a semi-analytical approach for analysing the relation between the galaxy size, halo and stellar mass in the massive  ($\rm{log(M_{star}/M_{\odot}) >11}$) satellites. 
They found that environments do not affect the compact sizes of satellites after their infall into the cluster. 

We found that red nuggets are located in all types of environments, and the fraction-$\delta$ relation is negative{, but constant within $1\sigma$. Also}, the fraction in LD ($33\pm7\%$) and HD ($22\pm6\%$) environments are consistent within $1\sigma$. 
However, note that \textit{red nuggets} do not survive in extreme LD environments (see the first panel in Fig.~\ref{fig:density_control_histo}, where only the most massive ($\rm{log(M_{star}/M_{\odot})>11.3}$) and the largest galaxies ($\rm{R_{e}}>2.5$ kpc) are found in the left tail of $\rm{log(1+\delta)}$ distribution). 
{We also found that the environment of \textit{red nuggets} is correlated with their stellar ages suggesting that \textit{relics} are found preferably in high-$\delta$ regions, while \textit{quiescent galaxies} are overabundant in low-$\delta$ environments (see Sec.~\ref{sec:rednugget_ages}).}

To explain the existence of red nuggets across such a wide variety of environments~\cite{Buitrago2018} proposed two channels that lead to red nuggets surviving to the local Universe. 
Namely, red nuggets in HD environments are red massive compact galaxies included in clusters at high redshift where a) high-velocity clusters prevent them from being cannibalised and/or b) the envelopes of the normal-size galaxies were removed through tidal interactions in clusters. 
The second channel leading to the {survival} of red nuggets in LD environments represents isolated red massive compact galaxies formed at high redshift where local densities were insufficient to experience merger events~\citep[e.g.][]{Aarseth1980}. {Though, only the most massive and the largest galaxies are able to survive in the extreme LD regions.}   

\begin{figure}
	\centerline{\includegraphics[width=0.49\textwidth]{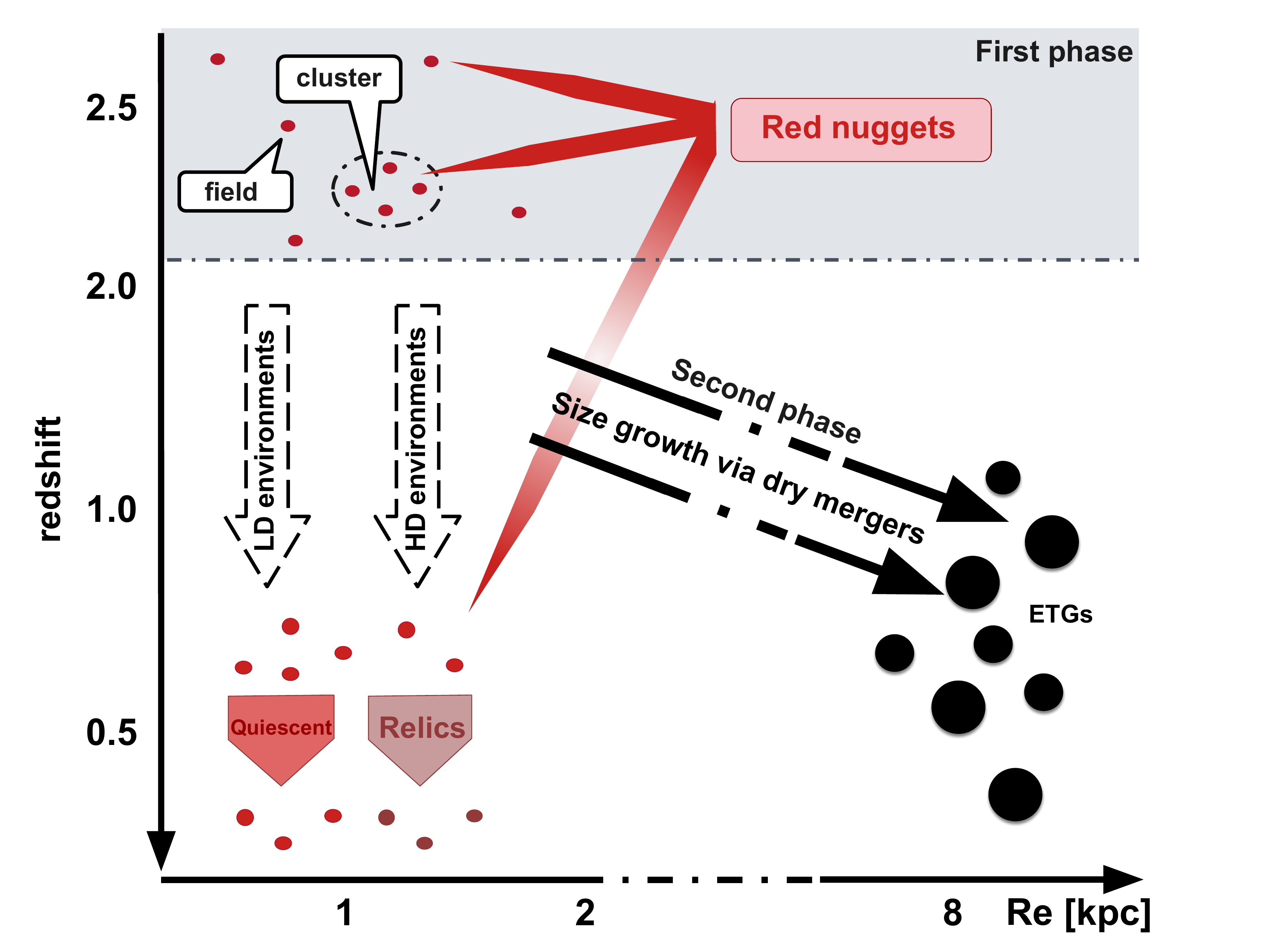}}	
	\caption{Schematic formation scenario for red nuggets. Red nuggets, massive, ultra-compact galaxies are formed at high redshift $\rm{\sim2-3}$ (first phase)
	and grow in size to form ETGs via dry mergers (second phase). Those red nuggets that did not grow into ETGs can be found in the local Universe as relics. The survivors of red nuggets can be found in different types of environments. They are protected from merger activities by the high relative velocities of galaxies within clusters and survive in LD environments where local densities are not sufficient to experience merger processes.}
	\label{fig:scenario}
\end{figure}

The two-channel scenario could explain observations of both low-redshift ($\rm{z<0.11}$) volume-limited (targeting clusters) observations of red nuggets at the centre of clusters~\citep[e.g.][]{poggianti2013} with red nuggets found in all types of environments drawn from high-completeness samples over the large effective area (180 $\rm{deg^2}$) at higher redshift~\citep[$\rm{0.15<z<0.5}$, e.g.][]{Buitrago2018,Tortora2020}.  
Consequently, extending the observations to intermediate redshift, our results fit the two-channel scenario, where the fraction of red nuggets in LD does not change significantly with redshift. 
Though, their fraction might slightly decrease as time passes, as red nuggets may experience merger activity (where local densities are sufficient) or grow disks around the bulges~\citep{Driver2013} 
and thus, not be observed in the local Universe. 
{At the same time, the fraction of red nuggets in HD environments may drop at higher redshifts ($\rm{z>1}$) where many of these cluster galaxies will have previously been within groups where mergers and low-velocity interactions were frequent. } 
In the local Universe, the fraction of red nuggets in HD environments is preserved as high-velocity clusters prevent them from merging or even increases due to the addition of newly normal-size red galaxies from which the envelopes were removed as time passes. 
The schematic overview of the two-channel scenario is presented in Fig.~\ref{fig:scenario}.

A similar mild negative trend (constant within $\rm{\lesssim1\sigma}$) of the fraction-$\delta$ relation as for \textit{red nuggets} has also been found for red, lower-mass compact galaxies (\textit{red intermediate-mass sample}, see the last panel in Fig.~\ref{fig:fraction}), revealing that \textit{red nuggets} are found in the same environments as the general population of red {ultra-}compact galaxies. 
This could be still explained under a two-channel scenario which is general for compact red galaxies, irrespective of their $\rm{M_{star}}$.

\section{Summary}\label{sec:summary}

In this work, we studied 42 red nuggets (with $\rm{log(M_{star}/M_{\odot} \geq 10.9)}$ and $\rm{R_{e}<1.5\mbox{ } kpc}$) at redshift $\rm{0.5<z<0.9}$ drawn from the VIPERS survey. 
We compared their environments, defined by local density contrast, to four control samples of red galaxies. 
The consistency of the number of \textit{red nuggets} residing in HD (10) and LD (11)  environments suggests that intermediate-redshift \textit{red nuggets} do not show preferences to HD environments as speculated~\citep[e.g.][]{poggianti2013, Peralta2016}, but are found in different environments as shown by other works~\citep[e.g.][]{Buitrago2018,Tortora2020}. 
{However, we found that the \textit{quiescent red nuggets} (of stellar ages $3-5$ Gyr) are overabundant in low-$\delta$ environments, while older ones (\textit{relic galaxies} with stellar ages $\geq$ 5 Gyr) are preferably found in high-$\delta$ regions, with \textit{recently quenched galaxies} with stellar ages $\leq 3$ Gyr placed both in low- and high-$\delta$ regions. 
The fractional decrease in the number of \textit{red nuggets} residing in HD with respect to LD regions is consistent with a flat distribution within $1\sigma$. }
Such a trend is in contradiction to the trends found for \textit{red galaxies} where the red fraction increases by 26$\pm11$\% from LD to HD environments (see Tab.~\ref{table:number_in_HD_LD} and \citealp{Siudek2022}). 
Our results show that the increasing fraction of red galaxies with density is driven by normal-size red galaxies (\textit{red normal-size sample}) {and compact-size red galaxies ($\rm{1.5<R_{e}<2.5\mbox{ } kpc}$)}, which are overabundant in HD environments. 
On the other hand, \textit{red nuggets} are found to reside in similar environments as red intermediate-mass ($\rm{log(M_{star}/M_{\odot})\sim10.64}$) ultra-compact galaxies ($\rm{R_{e}<1.5\mbox{ } kpc}$). 
The fraction of ultra-compact galaxies, regardless of their stellar mass, decreases when moving to HD environments and it is consistent within $\sim1\sigma$ suggesting that survival of ultra-compact galaxies is penalised in HD environments. 
Our results support the two-channel scenario, where red nuggets have survived to lower redshifts both in LD and HD environments. 
In LD environments their number do not change with redshift due to the scarcity of merger events, while in HD environments their number drops with redshift due to merger activity, but is preserved at low redshifts as the high velocities of clusters protect the surviving red nuggets from experiencing merger events.

\section*{Acknowledgements}
{The authors would like to thank the anonymous referee for the careful reading of the manuscript and the very useful comments. }
We want to also thank Angela Iovino and Ben Granett for their useful and constructive comments, 
Crescenzo Tortora for sharing data of his low-redshift results, and Mar Mezcua for the inspiration with schematic formation scenario. 
This work has been supported by the Polish National Agency for Academic Exchange (Bekker grant BPN/BEK/2021/1/00298/DEC/1), the European Union's Horizon 2020 Research and Innovation programme under the Maria Sklodowska-Curie grant agreement (No. 754510), and the Spanish Ministry of Science and Innovation through the Juan de la Cierva-formacion programme (FJC2018-038792-I). 
KL, KM and DD are grateful for support from the Polish National Science Centre via grant UMO-2018/30/E/ST9/00082 and UMO-2020/39/D/ST9/0072. AK acknowledges support from the First TEAM grant of the Foundation for Polish Science No. POIR.04.04.00-00-5D21/18-00 and the Polish National Agency for Academic Exchange grant No. BPN/BEK/2021/1/00319/DEC/1. AP and JK acknowledge the Polish National Science Centre (UMO-2018/30/M/ST9/0075) and Polish Ministry of Science and Higher Education grant DIR/WK/2018/12. CPH acknowledges support from ANID through Fondecyt Regular 2021 project no. 1211909.

\section*{Data Availability}
The VIPERS red nugget catalogue is publicly available and reported in~\citetalias{Lisiecki2022}. 
The environmental data underlying this paper will be shared upon request
to the corresponding author. 



\bibliographystyle{mnras}
\bibliography{VIPERSRedNuggets} 

\begin{thebibliography}{}
\makeatletter
\relax
\def\mn@urlcharsother{\let\do\@makeother \do\$\do\&\do\#\do\^\do\_\do\%\do\~}
\def\mn@doi{\begingroup\mn@urlcharsother \@ifnextchar [ {\mn@doi@}
  {\mn@doi@[]}}
\def\mn@doi@[#1]#2{\def\@tempa{#1}\ifx\@tempa\@empty \href
  {http://dx.doi.org/#2} {doi:#2}\else \href {http://dx.doi.org/#2} {#1}\fi
  \endgroup}
\def\mn@eprint#1#2{\mn@eprint@#1:#2::\@nil}
\def\mn@eprint@arXiv#1{\href {http://arxiv.org/abs/#1} {{\tt arXiv:#1}}}
\def\mn@eprint@dblp#1{\href {http://dblp.uni-trier.de/rec/bibtex/#1.xml}
  {dblp:#1}}
\def\mn@eprint@#1:#2:#3:#4\@nil{\def\@tempa {#1}\def\@tempb {#2}\def\@tempc
  {#3}\ifx \@tempc \@empty \let \@tempc \@tempb \let \@tempb \@tempa \fi \ifx
  \@tempb \@empty \def\@tempb {arXiv}\fi \@ifundefined
  {mn@eprint@\@tempb}{\@tempb:\@tempc}{\expandafter \expandafter \csname
  mn@eprint@\@tempb\endcsname \expandafter{\@tempc}}}

\bibitem[\protect\citeauthoryear{{Aarseth} \& {Fall}}{{Aarseth} \&
  {Fall}}{1980}]{Aarseth1980}
{Aarseth} S.~J.,  {Fall} S.~M.,  1980, \mn@doi [\apj] {10.1086/157716}, \href
  {https://ui.adsabs.harvard.edu/abs/1980ApJ...236...43A} {236, 43}

\bibitem[\protect\citeauthoryear{{Andreon}}{{Andreon}}{2020}]{Andreon2020}
{Andreon} S.,  2020, \mn@doi [\aap] {10.1051/0004-6361/202037584}, \href
  {https://ui.adsabs.harvard.edu/abs/2020A&A...640A..34A} {640, A34}

\bibitem[\protect\citeauthoryear{{Arnouts} et~al.,}{{Arnouts}
  et~al.}{2013}]{Arnouts2013}
{Arnouts} S.,  et~al., 2013, \mn@doi [\aap] {10.1051/0004-6361/201321768},
  \href {http://adsabs.harvard.edu/abs/2013A%26A...558A..67A} {558, A67}

\bibitem[\protect\citeauthoryear{{Baldry}, {Sullivan}, {Rani}  \&
  {Turner}}{{Baldry} et~al.}{2021}]{Baldry2021}
{Baldry} I.~K.,  {Sullivan} T.,  {Rani} R.,   {Turner} S.,  2021, \mn@doi
  [\mnras] {10.1093/mnras/staa3327}, \href
  {https://ui.adsabs.harvard.edu/abs/2021MNRAS.500.1557B} {500, 1557}

\bibitem[\protect\citeauthoryear{{Balogh}, {Morris}, {Yee}, {Carlberg}  \&
  {Ellingson}}{{Balogh} et~al.}{1997}]{Balogh1997}
{Balogh} M.~L.,  {Morris} S.~L.,  {Yee} H.~K.~C.,  {Carlberg} R.~G.,
  {Ellingson} E.,  1997, \mn@doi [\apjl] {10.1086/310927}, \href
  {https://ui.adsabs.harvard.edu/abs/1997ApJ...488L..75B} {488, L75}

\bibitem[\protect\citeauthoryear{{Balogh}, {Morris}, {Yee}, {Carlberg}  \&
  {Ellingson}}{{Balogh} et~al.}{1999}]{Balogh1999}
{Balogh} M.~L.,  {Morris} S.~L.,  {Yee} H.~K.~C.,  {Carlberg} R.~G.,
  {Ellingson} E.,  1999, \mn@doi [\apj] {10.1086/308056}, \href
  {https://ui.adsabs.harvard.edu/abs/1999ApJ...527...54B} {527, 54}

\bibitem[\protect\citeauthoryear{{Barro} et~al.,}{{Barro}
  et~al.}{2013}]{Barro2013}
{Barro} G.,  et~al., 2013, \mn@doi [\apj] {10.1088/0004-637X/765/2/104}, \href
  {https://ui.adsabs.harvard.edu/abs/2013ApJ...765..104B} {765, 104}

\bibitem[\protect\citeauthoryear{{Bertin} \& {Arnouts}}{{Bertin} \&
  {Arnouts}}{1996}]{Bertin1996}
{Bertin} E.,  {Arnouts} S.,  1996, \mn@doi [\aaps] {10.1051/aas:1996164}, \href
  {https://ui.adsabs.harvard.edu/abs/1996A&AS..117..393B} {117, 393}

\bibitem[\protect\citeauthoryear{{Boquien}, {Burgarella}, {Roehlly}, {Buat},
  {Ciesla}, {Corre}, {Inoue}  \& {Salas}}{{Boquien} et~al.}{2019}]{Boquien2019}
{Boquien} M.,  {Burgarella} D.,  {Roehlly} Y.,  {Buat} V.,  {Ciesla} L.,
  {Corre} D.,  {Inoue} A.~K.,   {Salas} H.,  2019, \mn@doi [\aap]
  {10.1051/0004-6361/201834156}, \href
  {https://ui.adsabs.harvard.edu/abs/2019A&A...622A.103B} {622, A103}

\bibitem[\protect\citeauthoryear{{Bravo}, {Robotham}, {Lagos}, {Davies},
  {Bellstedt}  \& {Thorne}}{{Bravo} et~al.}{2023}]{Bravo2023}
{Bravo} M.,  {Robotham} A. S.~G.,  {Lagos} C. d.~P.,  {Davies} L. J.~M.,
  {Bellstedt} S.,   {Thorne} J.~E.,  2023, \mn@doi [\mnras]
  {10.1093/mnras/stad1234}, \href
  {https://ui.adsabs.harvard.edu/abs/2023MNRAS.522.4481B} {522, 4481}

\bibitem[\protect\citeauthoryear{{Bruzual} \& {Charlot}}{{Bruzual} \&
  {Charlot}}{2003}]{Bruzal2003}
{Bruzual} G.,  {Charlot} S.,  2003, \mn@doi [\mnras]
  {10.1046/j.1365-8711.2003.06897.x}, \href
  {https://ui.adsabs.harvard.edu/abs/2003MNRAS.344.1000B} {344, 1000}

\bibitem[\protect\citeauthoryear{{Buitrago} et~al.,}{{Buitrago}
  et~al.}{2018}]{Buitrago2018}
{Buitrago} F.,  et~al., 2018, \mn@doi [\aap] {10.1051/0004-6361/201833785},
  \href {https://ui.adsabs.harvard.edu/abs/2018A&A...619A.137B} {619, A137}

\bibitem[\protect\citeauthoryear{{Calzetti}, {Armus}, {Bohlin}, {Kinney},
  {Koornneef}  \& {Storchi-Bergmann}}{{Calzetti} et~al.}{2000}]{Calzetti2000}
{Calzetti} D.,  {Armus} L.,  {Bohlin} R.~C.,  {Kinney} A.~L.,  {Koornneef} J.,
   {Storchi-Bergmann} T.,  2000, \mn@doi [\apj] {10.1086/308692}, \href
  {https://ui.adsabs.harvard.edu/abs/2000ApJ...533..682C} {533, 682}

\bibitem[\protect\citeauthoryear{{Capak}, {Abraham}, {Ellis}, {Mobasher},
  {Scoville}, {Sheth}  \& {Koekemoer}}{{Capak} et~al.}{2007}]{Capak2007}
{Capak} P.,  {Abraham} R.~G.,  {Ellis} R.~S.,  {Mobasher} B.,  {Scoville} N.,
  {Sheth} K.,   {Koekemoer} A.,  2007, \mn@doi [\apjs] {10.1086/518424}, \href
  {https://ui.adsabs.harvard.edu/abs/2007ApJS..172..284C} {172, 284}

\bibitem[\protect\citeauthoryear{{Cappellari}}{{Cappellari}}{2013}]{Cappellari2013}
{Cappellari} M.,  2013, \mn@doi [\apjl] {10.1088/2041-8205/778/1/L2}, \href
  {https://ui.adsabs.harvard.edu/abs/2013ApJ...778L...2C} {778, L2}

\bibitem[\protect\citeauthoryear{{Chabrier}}{{Chabrier}}{2003}]{Chabrier2003}
{Chabrier} G.,  2003, \mn@doi [\pasp] {10.1086/376392}, \href
  {https://ui.adsabs.harvard.edu/abs/2003PASP..115..763C} {115, 763}

\bibitem[\protect\citeauthoryear{{Chuter} et~al.,}{{Chuter}
  et~al.}{2011}]{Chuter2011}
{Chuter} R.~W.,  et~al., 2011, \mn@doi [\mnras]
  {10.1111/j.1365-2966.2011.18241.x}, \href
  {https://ui.adsabs.harvard.edu/abs/2011MNRAS.413.1678C} {413, 1678}

\bibitem[\protect\citeauthoryear{{Conselice} et~al.,}{{Conselice}
  et~al.}{2011}]{Conselice2011}
{Conselice} C.~J.,  et~al., 2011, \mn@doi [\mnras]
  {10.1111/j.1365-2966.2010.18113.x}, \href
  {https://ui.adsabs.harvard.edu/abs/2011MNRAS.413...80C} {413, 80}

\bibitem[\protect\citeauthoryear{{Cooper} et~al.,}{{Cooper}
  et~al.}{2007}]{Cooper2007}
{Cooper} M.~C.,  et~al., 2007, \mn@doi [\mnras]
  {10.1111/j.1365-2966.2007.11534.x}, \href
  {https://ui.adsabs.harvard.edu/abs/2007MNRAS.376.1445C} {376, 1445}

\bibitem[\protect\citeauthoryear{{Cooper} et~al.,}{{Cooper}
  et~al.}{2012}]{Cooper2012}
{Cooper} M.~C.,  et~al., 2012, \mn@doi [\mnras]
  {10.1111/j.1365-2966.2011.19938.x}, \href
  {https://ui.adsabs.harvard.edu/abs/2012MNRAS.419.3018C} {419, 3018}

\bibitem[\protect\citeauthoryear{{Cucciati} et~al.,}{{Cucciati}
  et~al.}{2014}]{Cucciati2014}
{Cucciati} O.,  et~al., 2014, \mn@doi [\aap] {10.1051/0004-6361/201423409},
  \href {https://ui.adsabs.harvard.edu/abs/2014A&A...565A..67C} {565, A67}

\bibitem[\protect\citeauthoryear{{Cucciati} et~al.,}{{Cucciati}
  et~al.}{2017}]{Cucciati2017}
{Cucciati} O.,  et~al., 2017, \mn@doi [\aap] {10.1051/0004-6361/201630113},
  \href {https://ui.adsabs.harvard.edu/abs/2017A&A...602A..15C} {602, A15}

\bibitem[\protect\citeauthoryear{{Daddi} et~al.,}{{Daddi}
  et~al.}{2005}]{Daddi2005}
{Daddi} E.,  et~al., 2005, \mn@doi [\apj] {10.1086/430104}, \href
  {https://ui.adsabs.harvard.edu/abs/2005ApJ...626..680D} {626, 680}

\bibitem[\protect\citeauthoryear{{Dale}, {Helou}, {Magdis}, {Armus},
  {D{\'\i}az-Santos}  \& {Shi}}{{Dale} et~al.}{2014}]{dale2014}
{Dale} D.~A.,  {Helou} G.,  {Magdis} G.~E.,  {Armus} L.,  {D{\'\i}az-Santos}
  T.,   {Shi} Y.,  2014, \mn@doi [\apj] {10.1088/0004-637X/784/1/83}, \href
  {https://ui.adsabs.harvard.edu/abs/2014ApJ...784...83D} {784, 83}

\bibitem[\protect\citeauthoryear{{Damjanov} et~al.,}{{Damjanov}
  et~al.}{2009}]{Damjanov2009}
{Damjanov} I.,  et~al., 2009, \mn@doi [\apj] {10.1088/0004-637X/695/1/101},
  \href {https://ui.adsabs.harvard.edu/abs/2009ApJ...695..101D} {695, 101}

\bibitem[\protect\citeauthoryear{{Damjanov}, {Zahid}, {Geller}  \&
  {Hwang}}{{Damjanov} et~al.}{2015}]{Damjanov2015}
{Damjanov} I.,  {Zahid} H.~J.,  {Geller} M.~J.,   {Hwang} H.~S.,  2015, \mn@doi
  [\apj] {10.1088/0004-637X/815/2/104}, \href
  {https://ui.adsabs.harvard.edu/abs/2015ApJ...815..104D} {815, 104}

\bibitem[\protect\citeauthoryear{{Davidzon} et~al.,}{{Davidzon}
  et~al.}{2016}]{Davidzon2016}
{Davidzon} I.,  et~al., 2016, \mn@doi [\aap] {10.1051/0004-6361/201527129},
  \href {https://ui.adsabs.harvard.edu/abs/2016A&A...586A..23D} {586, A23}

\bibitem[\protect\citeauthoryear{{Davidzon} et~al.,}{{Davidzon}
  et~al.}{2017}]{Davidzon2017}
{Davidzon} I.,  et~al., 2017, \mn@doi [\aap] {10.1051/0004-6361/201730419},
  \href {https://ui.adsabs.harvard.edu/abs/2017A&A...605A..70D} {605, A70}

\bibitem[\protect\citeauthoryear{{Dekel}, {Sari}  \& {Ceverino}}{{Dekel}
  et~al.}{2009}]{Dekel2009}
{Dekel} A.,  {Sari} R.,   {Ceverino} D.,  2009, \mn@doi [\apj]
  {10.1088/0004-637X/703/1/785}, \href
  {https://ui.adsabs.harvard.edu/abs/2009ApJ...703..785D} {703, 785}

\bibitem[\protect\citeauthoryear{{Di Matteo}, {Springel}  \& {Hernquist}}{{Di
  Matteo} et~al.}{2005}]{DiMatteo2005}
{Di Matteo} T.,  {Springel} V.,   {Hernquist} L.,  2005, \mn@doi [\nat]
  {10.1038/nature03335}, \href
  {https://ui.adsabs.harvard.edu/abs/2005Natur.433..604D} {433, 604}

\bibitem[\protect\citeauthoryear{{Dressler}}{{Dressler}}{1980}]{Dressler1980}
{Dressler} A.,  1980, \mn@doi [\apj] {10.1086/157753}, \href
  {https://ui.adsabs.harvard.edu/abs/1980ApJ...236..351D} {236, 351}

\bibitem[\protect\citeauthoryear{{Dressler} et~al.,}{{Dressler}
  et~al.}{1997}]{Dressler1997}
{Dressler} A.,  et~al., 1997, \mn@doi [\apj] {10.1086/304890}, \href
  {https://ui.adsabs.harvard.edu/abs/1997ApJ...490..577D} {490, 577}

\bibitem[\protect\citeauthoryear{{Driver}, {Robotham}, {Bland-Hawthorn},
  {Brown}, {Hopkins}, {Liske}, {Phillipps}  \& {Wilkins}}{{Driver}
  et~al.}{2013}]{Driver2013}
{Driver} S.~P.,  {Robotham} A.~S.~G.,  {Bland-Hawthorn} J.,  {Brown} M.,
  {Hopkins} A.,  {Liske} J.,  {Phillipps} S.,   {Wilkins} S.,  2013, \mn@doi
  [\mnras] {10.1093/mnras/sts717}, \href
  {https://ui.adsabs.harvard.edu/abs/2013MNRAS.430.2622D} {430, 2622}

\bibitem[\protect\citeauthoryear{{Figueira} et~al.,}{{Figueira}
  et~al.}{2022}]{Figueira2022}
{Figueira} M.,  et~al., 2022, \mn@doi [\aap] {10.1051/0004-6361/202141701},
  \href {https://ui.adsabs.harvard.edu/abs/2022A&A...667A..29F} {667, A29}

\bibitem[\protect\citeauthoryear{{Flores-Freitas}, {Chies-Santos},
  {Furlanetto}, {De Rossi}, {Ferreira}, {Zenocratti}  \&
  {Alamo-Mart{\'\i}nez}}{{Flores-Freitas} et~al.}{2022}]{Flores2022}
{Flores-Freitas} R.,  {Chies-Santos} A.~L.,  {Furlanetto} C.,  {De Rossi}
  M.~E.,  {Ferreira} L.,  {Zenocratti} L.~J.,   {Alamo-Mart{\'\i}nez} K.~A.,
  2022, \mn@doi [\mnras] {10.1093/mnras/stac187}, \href
  {https://ui.adsabs.harvard.edu/abs/2022MNRAS.512..245F} {512, 245}

\bibitem[\protect\citeauthoryear{{Fontana} et~al.,}{{Fontana}
  et~al.}{2006}]{Fontana2006}
{Fontana} A.,  et~al., 2006, \mn@doi [\aap] {10.1051/0004-6361:20065475}, \href
  {https://ui.adsabs.harvard.edu/abs/2006A&A...459..745F} {459, 745}

\bibitem[\protect\citeauthoryear{{Gallazzi}, {Charlot}, {Brinchmann}  \&
  {White}}{{Gallazzi} et~al.}{2006}]{Gallazzi2006}
{Gallazzi} A.,  {Charlot} S.,  {Brinchmann} J.,   {White} S. D.~M.,  2006,
  \mn@doi [\mnras] {10.1111/j.1365-2966.2006.10548.x}, \href
  {https://ui.adsabs.harvard.edu/abs/2006MNRAS.370.1106G} {370, 1106}

\bibitem[\protect\citeauthoryear{{Gallazzi}, {Bell}, {Zibetti}, {Brinchmann}
  \& {Kelson}}{{Gallazzi} et~al.}{2014}]{Gallazzi2014}
{Gallazzi} A.,  {Bell} E.~F.,  {Zibetti} S.,  {Brinchmann} J.,   {Kelson}
  D.~D.,  2014, \mn@doi [\apj] {10.1088/0004-637X/788/1/72}, \href
  {https://ui.adsabs.harvard.edu/abs/2014ApJ...788...72G} {788, 72}

\bibitem[\protect\citeauthoryear{{Gargiulo} et~al.,}{{Gargiulo}
  et~al.}{2017}]{Gargiulo2017}
{Gargiulo} A.,  et~al., 2017, \mn@doi [\aap] {10.1051/0004-6361/201630112},
  \href {https://ui.adsabs.harvard.edu/abs/2017A&A...606A.113G} {606, A113}

\bibitem[\protect\citeauthoryear{{Gargiulo} et~al.,}{{Gargiulo}
  et~al.}{2019}]{Gargiulo2019}
{Gargiulo} A.,  et~al., 2019, \mn@doi [\aap] {10.1051/0004-6361/201833600},
  \href {https://ui.adsabs.harvard.edu/abs/2019A&A...631A..15G} {631, A15}

\bibitem[\protect\citeauthoryear{{Garilli} et~al.,}{{Garilli}
  et~al.}{2014}]{Garilli2014}
{Garilli} B.,  et~al., 2014, \mn@doi [\aap] {10.1051/0004-6361/201322790},
  \href {http://adsabs.harvard.edu/abs/2014A%26A...562A..23G} {562, A23}

\bibitem[\protect\citeauthoryear{{Goranova, Y.} et~al.,}{{Goranova, Y.}
  et~al.}{2009}]{goranowa09}
{Goranova, Y.} et~al., 2009, Technical report, The CFHTLS T0006 Release.
Terapix/Institut d’Astrophysique de Paris

\bibitem[\protect\citeauthoryear{{Guglielmo}, {Poggianti}, {Moretti}, {Fritz},
  {Calvi}, {Vulcani}, {Fasano}  \& {Paccagnella}}{{Guglielmo}
  et~al.}{2015}]{Guglielmo2015}
{Guglielmo} V.,  {Poggianti} B.~M.,  {Moretti} A.,  {Fritz} J.,  {Calvi} R.,
  {Vulcani} B.,  {Fasano} G.,   {Paccagnella} A.,  2015, \mn@doi [\mnras]
  {10.1093/mnras/stv757}, \href
  {https://ui.adsabs.harvard.edu/abs/2015MNRAS.450.2749G} {450, 2749}

\bibitem[\protect\citeauthoryear{{Guzzo} et~al.,}{{Guzzo}
  et~al.}{2014}]{Guzzo2014}
{Guzzo} L.,  et~al., 2014, \mn@doi [\aap] {10.1051/0004-6361/201321489}, \href
  {http://adsabs.harvard.edu/abs/2014A%26A...566A.108G} {566, A108}

\bibitem[\protect\citeauthoryear{{Haines} et~al.,}{{Haines}
  et~al.}{2017}]{Haines2017}
{Haines} C.~P.,  et~al., 2017, \mn@doi [\aap] {10.1051/0004-6361/201630118},
  \href {https://ui.adsabs.harvard.edu/abs/2017A&A...605A...4H} {605, A4}

\bibitem[\protect\citeauthoryear{{Hilz}, {Naab}  \& {Ostriker}}{{Hilz}
  et~al.}{2013}]{Hilz2013}
{Hilz} M.,  {Naab} T.,   {Ostriker} J.~P.,  2013, \mn@doi [\mnras]
  {10.1093/mnras/sts501}, \href
  {https://ui.adsabs.harvard.edu/abs/2013MNRAS.429.2924H} {429, 2924}

\bibitem[\protect\citeauthoryear{{Hogg} et~al.,}{{Hogg}
  et~al.}{2003}]{Hogg2003}
{Hogg} D.~W.,  et~al., 2003, \mn@doi [\apjl] {10.1086/374238}, \href
  {https://ui.adsabs.harvard.edu/abs/2003ApJ...585L...5H} {585, L5}

\bibitem[\protect\citeauthoryear{{Hopkins}, {Hernquist}, {Cox}, {Di Matteo},
  {Robertson}  \& {Springel}}{{Hopkins} et~al.}{2006}]{Hopkins2006}
{Hopkins} P.~F.,  {Hernquist} L.,  {Cox} T.~J.,  {Di Matteo} T.,  {Robertson}
  B.,   {Springel} V.,  2006, \mn@doi [\apjs] {10.1086/499298}, \href
  {https://ui.adsabs.harvard.edu/abs/2006ApJS..163....1H} {163, 1}

\bibitem[\protect\citeauthoryear{{Hopkins}, {Hernquist}, {Cox}  \&
  {Kere{\v{s}}}}{{Hopkins} et~al.}{2008}]{Hopkins2008}
{Hopkins} P.~F.,  {Hernquist} L.,  {Cox} T.~J.,   {Kere{\v{s}}} D.,  2008,
  \mn@doi [\apjs] {10.1086/524362}, \href
  {https://ui.adsabs.harvard.edu/abs/2008ApJS..175..356H} {175, 356}

\bibitem[\protect\citeauthoryear{{Hopkins}, {Hernquist}, {Cox}, {Keres}  \&
  {Wuyts}}{{Hopkins} et~al.}{2009}]{Hopkins2009}
{Hopkins} P.~F.,  {Hernquist} L.,  {Cox} T.~J.,  {Keres} D.,   {Wuyts} S.,
  2009, \mn@doi [\apj] {10.1088/0004-637X/691/2/1424}, \href
  {https://ui.adsabs.harvard.edu/abs/2009ApJ...691.1424H} {691, 1424}

\bibitem[\protect\citeauthoryear{{Huertas-Company} et~al.,}{{Huertas-Company}
  et~al.}{2013}]{HuertasCompany2013}
{Huertas-Company} M.,  et~al., 2013, \mn@doi [\mnras] {10.1093/mnras/sts150},
  \href {https://ui.adsabs.harvard.edu/abs/2013MNRAS.428.1715H} {428, 1715}

\bibitem[\protect\citeauthoryear{{Inoue}}{{Inoue}}{2011}]{Inoue2011}
{Inoue} A.~K.,  2011, \mn@doi [\mnras] {10.1111/j.1365-2966.2011.18906.x},
  \href {https://ui.adsabs.harvard.edu/abs/2011MNRAS.415.2920I} {415, 2920}

\bibitem[\protect\citeauthoryear{{Jarvis} et~al.,}{{Jarvis}
  et~al.}{2013}]{jarvis13}
{Jarvis} M.~J.,  et~al., 2013, \mn@doi [\mnras] {10.1093/mnras/sts118}, \href
  {https://ui.adsabs.harvard.edu/abs/2013MNRAS.428.1281J} {428, 1281}

\bibitem[\protect\citeauthoryear{{Kauffmann} et~al.,}{{Kauffmann}
  et~al.}{2003}]{Kauffmann2003}
{Kauffmann} G.,  et~al., 2003, \mn@doi [\mnras]
  {10.1046/j.1365-8711.2003.06291.x}, \href
  {https://ui.adsabs.harvard.edu/abs/2003MNRAS.341...33K} {341, 33}

\bibitem[\protect\citeauthoryear{{Kauffmann}, {White}, {Heckman}, {M{\'e}nard},
  {Brinchmann}, {Charlot}, {Tremonti}  \& {Brinkmann}}{{Kauffmann}
  et~al.}{2004}]{Kauffmann2004}
{Kauffmann} G.,  {White} S. D.~M.,  {Heckman} T.~M.,  {M{\'e}nard} B.,
  {Brinchmann} J.,  {Charlot} S.,  {Tremonti} C.,   {Brinkmann} J.,  2004,
  \mn@doi [\mnras] {10.1111/j.1365-2966.2004.08117.x}, \href
  {https://ui.adsabs.harvard.edu/abs/2004MNRAS.353..713K} {353, 713}

\bibitem[\protect\citeauthoryear{{Kelkar}, {Arag{\'o}n-Salamanca}, {Gray},
  {Maltby}, {Vulcani}, {De Lucia}, {Poggianti}  \& {Zaritsky}}{{Kelkar}
  et~al.}{2015}]{Kelkar2015}
{Kelkar} K.,  {Arag{\'o}n-Salamanca} A.,  {Gray} M.~E.,  {Maltby} D.,
  {Vulcani} B.,  {De Lucia} G.,  {Poggianti} B.~M.,   {Zaritsky} D.,  2015,
  \mn@doi [\mnras] {10.1093/mnras/stv670}, \href
  {https://ui.adsabs.harvard.edu/abs/2015MNRAS.450.1246K} {450, 1246}

\bibitem[\protect\citeauthoryear{{Kere{\v{s}}}, {Katz}, {Weinberg}  \&
  {Dav{\'e}}}{{Kere{\v{s}}} et~al.}{2005}]{Keres2005}
{Kere{\v{s}}} D.,  {Katz} N.,  {Weinberg} D.~H.,   {Dav{\'e}} R.,  2005,
  \mn@doi [\mnras] {10.1111/j.1365-2966.2005.09451.x}, \href
  {https://ui.adsabs.harvard.edu/abs/2005MNRAS.363....2K} {363, 2}

\bibitem[\protect\citeauthoryear{{Krywult} et~al.,}{{Krywult}
  et~al.}{2017}]{Krywult2017}
{Krywult} J.,  et~al., 2017, \mn@doi [\aap] {10.1051/0004-6361/201628953},
  \href {http://adsabs.harvard.edu/abs/2017A%26A...598A.120K} {598, A120}

\bibitem[\protect\citeauthoryear{{Lani} et~al.,}{{Lani}
  et~al.}{2013}]{Lani2013}
{Lani} C.,  et~al., 2013, \mn@doi [\mnras] {10.1093/mnras/stt1275}, \href
  {https://ui.adsabs.harvard.edu/abs/2013MNRAS.435..207L} {435, 207}

\bibitem[\protect\citeauthoryear{{Lapi} et~al.,}{{Lapi} et~al.}{2018}]{lapi18}
{Lapi} A.,  et~al., 2018, \mn@doi [\apj] {10.3847/1538-4357/aab6af}, \href
  {https://ui.adsabs.harvard.edu/abs/2018ApJ...857...22L} {857, 22}

\bibitem[\protect\citeauthoryear{{Lewis} et~al.,}{{Lewis}
  et~al.}{2002}]{Lewis2002}
{Lewis} I.,  et~al., 2002, \mn@doi [\mnras] {10.1046/j.1365-8711.2002.05558.x},
  \href {https://ui.adsabs.harvard.edu/abs/2002MNRAS.334..673L} {334, 673}

\bibitem[\protect\citeauthoryear{{Lisiecki}, {Ma{\l}ek}, {Siudek}, {Pollo},
  {Krywult}, {Karska}  \& {Junais}}{{Lisiecki} et~al.}{2023}]{Lisiecki2022}
{Lisiecki} K.,  {Ma{\l}ek} K.,  {Siudek} M.,  {Pollo} A.,  {Krywult} J.,
  {Karska} A.,   {Junais} 2023, \mn@doi [\aap] {10.1051/0004-6361/202243616},
  \href {https://ui.adsabs.harvard.edu/abs/2023A&A...669A..95L} {669, A95}

\bibitem[\protect\citeauthoryear{{Maltby} et~al.,}{{Maltby}
  et~al.}{2010}]{Maltby2010}
{Maltby} D.~T.,  et~al., 2010, \mn@doi [\mnras]
  {10.1111/j.1365-2966.2009.15953.x}, \href
  {https://ui.adsabs.harvard.edu/abs/2010MNRAS.402..282M} {402, 282}

\bibitem[\protect\citeauthoryear{{Martin} et~al.,}{{Martin}
  et~al.}{2005}]{Martin2005}
{Martin} D.~C.,  et~al., 2005, \mn@doi [\apjl] {10.1086/426387}, \href
  {https://ui.adsabs.harvard.edu/abs/2005ApJ...619L...1M} {619, L1}

\bibitem[\protect\citeauthoryear{{McLeod}, {McLure}, {Dunlop}, {Cullen},
  {Carnall}  \& {Duncan}}{{McLeod} et~al.}{2021}]{McLeod2021}
{McLeod} D.~J.,  {McLure} R.~J.,  {Dunlop} J.~S.,  {Cullen} F.,  {Carnall}
  A.~C.,   {Duncan} K.,  2021, \mn@doi [\mnras] {10.1093/mnras/stab731}, \href
  {https://ui.adsabs.harvard.edu/abs/2021MNRAS.503.4413M} {503, 4413}

\bibitem[\protect\citeauthoryear{{Mezcua}, {Siudek}, {Suh}, {Valiante},
  {Spinoso}  \& {Bonoli}}{{Mezcua} et~al.}{2023}]{Mezcua2023}
{Mezcua} M.,  {Siudek} M.,  {Suh} H.,  {Valiante} R.,  {Spinoso} D.,   {Bonoli}
  S.,  2023, \mn@doi [\apjl] {10.3847/2041-8213/acae25}, \href
  {https://ui.adsabs.harvard.edu/abs/2023ApJ...943L...5M} {943, L5}

\bibitem[\protect\citeauthoryear{{Moutard} et~al.,}{{Moutard}
  et~al.}{2016}]{moutard2016b}
{Moutard} T.,  et~al., 2016, \mn@doi [\aap] {10.1051/0004-6361/201527294},
  \href {http://adsabs.harvard.edu/abs/2016A\%26A...590A.103M} {590, A103}

\bibitem[\protect\citeauthoryear{{Mowla} et~al.,}{{Mowla}
  et~al.}{2019}]{Mowla2019}
{Mowla} L.~A.,  et~al., 2019, \mn@doi [\apj] {10.3847/1538-4357/ab290a}, \href
  {https://ui.adsabs.harvard.edu/abs/2019ApJ...880...57M} {880, 57}

\bibitem[\protect\citeauthoryear{{Naab}, {Khochfar}  \& {Burkert}}{{Naab}
  et~al.}{2006}]{Naab2006}
{Naab} T.,  {Khochfar} S.,   {Burkert} A.,  2006, \mn@doi [\apjl]
  {10.1086/500205}, \href
  {https://ui.adsabs.harvard.edu/abs/2006ApJ...636L..81N} {636, L81}

\bibitem[\protect\citeauthoryear{{Naab}, {Johansson}, {Ostriker}  \&
  {Efstathiou}}{{Naab} et~al.}{2007}]{Naab2007}
{Naab} T.,  {Johansson} P.~H.,  {Ostriker} J.~P.,   {Efstathiou} G.,  2007,
  \mn@doi [\apj] {10.1086/510841}, \href
  {https://ui.adsabs.harvard.edu/abs/2007ApJ...658..710N} {658, 710}

\bibitem[\protect\citeauthoryear{{Naab}, {Johansson}  \& {Ostriker}}{{Naab}
  et~al.}{2009}]{Naab2009}
{Naab} T.,  {Johansson} P.~H.,   {Ostriker} J.~P.,  2009, \mn@doi [\apjl]
  {10.1088/0004-637X/699/2/L178}, \href
  {https://ui.adsabs.harvard.edu/abs/2009ApJ...699L.178N} {699, L178}

\bibitem[\protect\citeauthoryear{{Naidu} et~al.,}{{Naidu}
  et~al.}{2022}]{Naidu2022}
{Naidu} R.~P.,  et~al., 2022, \mn@doi [\apjl] {10.3847/2041-8213/ac9b22}, \href
  {https://ui.adsabs.harvard.edu/abs/2022ApJ...940L..14N} {940, L14}

\bibitem[\protect\citeauthoryear{{Nair}, {van den Bergh}  \& {Abraham}}{{Nair}
  et~al.}{2010}]{Nair2010}
{Nair} P.~B.,  {van den Bergh} S.,   {Abraham} R.~G.,  2010, \mn@doi [\apj]
  {10.1088/0004-637X/715/1/606}, \href
  {https://ui.adsabs.harvard.edu/abs/2010ApJ...715..606N} {715, 606}

\bibitem[\protect\citeauthoryear{{Oser}, {Ostriker}, {Naab}, {Johansson}  \&
  {Burkert}}{{Oser} et~al.}{2010}]{Oser2010}
{Oser} L.,  {Ostriker} J.~P.,  {Naab} T.,  {Johansson} P.~H.,   {Burkert} A.,
  2010, \mn@doi [\apj] {10.1088/0004-637X/725/2/2312}, \href
  {https://ui.adsabs.harvard.edu/abs/2010ApJ...725.2312O} {725, 2312}

\bibitem[\protect\citeauthoryear{{Oser}, {Naab}, {Ostriker}  \&
  {Johansson}}{{Oser} et~al.}{2012}]{Oser2012}
{Oser} L.,  {Naab} T.,  {Ostriker} J.~P.,   {Johansson} P.~H.,  2012, \mn@doi
  [\apj] {10.1088/0004-637X/744/1/63}, \href
  {https://ui.adsabs.harvard.edu/abs/2012ApJ...744...63O} {744, 63}

\bibitem[\protect\citeauthoryear{{Papovich} et~al.,}{{Papovich}
  et~al.}{2012}]{Papovich2012}
{Papovich} C.,  et~al., 2012, \mn@doi [\apj] {10.1088/0004-637X/750/2/93},
  \href {https://ui.adsabs.harvard.edu/abs/2012ApJ...750...93P} {750, 93}

\bibitem[\protect\citeauthoryear{{Paulino-Afonso} et~al.,}{{Paulino-Afonso}
  et~al.}{2019}]{PaulinoAfonso2019}
{Paulino-Afonso} A.,  et~al., 2019, \mn@doi [\aap]
  {10.1051/0004-6361/201935137}, \href
  {https://ui.adsabs.harvard.edu/abs/2019A&A...630A..57P} {630, A57}

\bibitem[\protect\citeauthoryear{{Peng}, {Ho}, {Impey}  \& {Rix}}{{Peng}
  et~al.}{2002}]{Peng2002}
{Peng} C.~Y.,  {Ho} L.~C.,  {Impey} C.~D.,   {Rix} H.-W.,  2002, \mn@doi [\aj]
  {10.1086/340952}, \href
  {https://ui.adsabs.harvard.edu/abs/2002AJ....124..266P} {124, 266}

\bibitem[\protect\citeauthoryear{{Peng} et~al.,}{{Peng}
  et~al.}{2010}]{Peng2010}
{Peng} Y.-j.,  et~al., 2010, \mn@doi [\apj] {10.1088/0004-637X/721/1/193},
  \href {http://adsabs.harvard.edu/abs/2010ApJ...721..193P} {721, 193}

\bibitem[\protect\citeauthoryear{{Peralta de Arriba}, {Quilis}, {Trujillo},
  {Cebri{\'a}n}  \& {Balcells}}{{Peralta de Arriba} et~al.}{2016}]{Peralta2016}
{Peralta de Arriba} L.,  {Quilis} V.,  {Trujillo} I.,  {Cebri{\'a}n} M.,
  {Balcells} M.,  2016, \mn@doi [\mnras] {10.1093/mnras/stw1240}, \href
  {https://ui.adsabs.harvard.edu/abs/2016MNRAS.461..156P} {461, 156}

\bibitem[\protect\citeauthoryear{{Pierre} et~al.,}{{Pierre}
  et~al.}{2004}]{Pierre2004}
{Pierre} M.,  et~al., 2004, \mn@doi [\jcap] {10.1088/1475-7516/2004/09/011},
  \href {https://ui.adsabs.harvard.edu/abs/2004JCAP...09..011P} {2004, 011}

\bibitem[\protect\citeauthoryear{{Pistis} et~al.,}{{Pistis}
  et~al.}{2022}]{Pistis2022}
{Pistis} F.,  et~al., 2022, \mn@doi [\aap] {10.1051/0004-6361/202142430}, \href
  {https://ui.adsabs.harvard.edu/abs/2022A&A...663A.162P} {663, A162}

\bibitem[\protect\citeauthoryear{{Poggianti} et~al.,}{{Poggianti}
  et~al.}{2013}]{poggianti2013}
{Poggianti} B.~M.,  et~al., 2013, \mn@doi [\apj] {10.1088/0004-637X/762/2/77},
  \href {https://ui.adsabs.harvard.edu/abs/2013ApJ...762...77P} {762, 77}

\bibitem[\protect\citeauthoryear{{Puget} et~al.,}{{Puget}
  et~al.}{2004}]{Puget2004S}
{Puget} P.,  et~al., 2004, in {Moorwood} A. F.~M.,  {Iye} M.,  eds,  Society of
  Photo-Optical Instrumentation Engineers (SPIE) Conference Series Vol. 5492,
  Ground-based Instrumentation for Astronomy. pp 978--987,
  \mn@doi{10.1117/12.551097}

\bibitem[\protect\citeauthoryear{{Quilis} \& {Trujillo}}{{Quilis} \&
  {Trujillo}}{2013}]{Quilis2013}
{Quilis} V.,  {Trujillo} I.,  2013, \mn@doi [\apjl]
  {10.1088/2041-8205/773/1/L8}, \href
  {https://ui.adsabs.harvard.edu/abs/2013ApJ...773L...8Q} {773, L8}

\bibitem[\protect\citeauthoryear{{Renard} et~al.,}{{Renard}
  et~al.}{2022}]{Renard2022}
{Renard} P.,  et~al., 2022, \mn@doi [\mnras] {10.1093/mnras/stac1730}, \href
  {https://ui.adsabs.harvard.edu/abs/2022MNRAS.515..146R} {515, 146}

\bibitem[\protect\citeauthoryear{{Sazonova} et~al.,}{{Sazonova}
  et~al.}{2020}]{Sazonova2020}
{Sazonova} E.,  et~al., 2020, \mn@doi [\apj] {10.3847/1538-4357/aba42f}, \href
  {https://ui.adsabs.harvard.edu/abs/2020ApJ...899...85S} {899, 85}

\bibitem[\protect\citeauthoryear{{Schmidt}}{{Schmidt}}{1968}]{Schmidt1968}
{Schmidt} M.,  1968, \mn@doi [\apj] {10.1086/149446}, \href
  {https://ui.adsabs.harvard.edu/abs/1968ApJ...151..393S} {151, 393}

\bibitem[\protect\citeauthoryear{{Schnorr-M{\"u}ller}
  et~al.,}{{Schnorr-M{\"u}ller} et~al.}{2021}]{schnorr22}
{Schnorr-M{\"u}ller} A.,  et~al., 2021, \mn@doi [\mnras]
  {10.1093/mnras/stab2116}, \href
  {https://ui.adsabs.harvard.edu/abs/2021MNRAS.507..300S} {507, 300}

\bibitem[\protect\citeauthoryear{{Scodeggio} et~al.,}{{Scodeggio}
  et~al.}{2018}]{Scodeggio2018}
{Scodeggio} M.,  et~al., 2018, \mn@doi [\aap] {10.1051/0004-6361/201630114},
  \href {http://adsabs.harvard.edu/abs/2018A%26A...609A..84S} {609, A84}

\bibitem[\protect\citeauthoryear{{Scognamiglio} et~al.,}{{Scognamiglio}
  et~al.}{2020}]{Scognamiglio2020}
{Scognamiglio} D.,  et~al., 2020, \mn@doi [\apj] {10.3847/1538-4357/ab7db3},
  \href {https://ui.adsabs.harvard.edu/abs/2020ApJ...893....4S} {893, 4}

\bibitem[\protect\citeauthoryear{{Siudek} et~al.,}{{Siudek}
  et~al.}{2017}]{siudek2017}
{Siudek} M.,  et~al., 2017, \mn@doi [\aap] {10.1051/0004-6361/201628951}, \href
  {http://adsabs.harvard.edu/abs/2017A%26A...597A.107S} {597, A107}

\bibitem[\protect\citeauthoryear{{Siudek} et~al.,}{{Siudek}
  et~al.}{2018a}]{Siudek2018b}
{Siudek} M.,  et~al., 2018a, arXiv e-prints, \href
  {https://ui.adsabs.harvard.edu/abs/2018arXiv180509905S} {p. arXiv:1805.09905}

\bibitem[\protect\citeauthoryear{{Siudek} et~al.,}{{Siudek}
  et~al.}{2018b}]{Siudek2018a}
{Siudek} M.,  et~al., 2018b, \mn@doi [\aap] {10.1051/0004-6361/201832784},
  \href {https://ui.adsabs.harvard.edu/abs/2018A&A...617A..70S} {617, A70}

\bibitem[\protect\citeauthoryear{{Siudek}, {Lisiecki}, {Mezcua}, {Ma{\l}ek},
  {Pollo}, {Krywult}, {Karska}  \& {Junais}}{{Siudek}
  et~al.}{2022a}]{Siudek2022Proceeding}
{Siudek} M.,  {Lisiecki} K.,  {Mezcua} M.,  {Ma{\l}ek} K.,  {Pollo} A.,
  {Krywult} J.,  {Karska} A.,   {Junais} 2022a, \mn@doi [arXiv e-prints]
  {10.48550/arXiv.2211.11792}, \href
  {https://ui.adsabs.harvard.edu/abs/2022arXiv221111792S} {p. arXiv:2211.11792}

\bibitem[\protect\citeauthoryear{{Siudek} et~al.,}{{Siudek}
  et~al.}{2022b}]{Siudek2022}
{Siudek} M.,  et~al., 2022b, \mn@doi [\aap] {10.1051/0004-6361/202243613},
  \href {https://ui.adsabs.harvard.edu/abs/2022A&A...666A.131S} {666, A131}

\bibitem[\protect\citeauthoryear{{Siudek}, {Mezcua}  \& {Krywult}}{{Siudek}
  et~al.}{2023}]{Siudek2023}
{Siudek} M.,  {Mezcua} M.,   {Krywult} J.,  2023, \mn@doi [\mnras]
  {10.1093/mnras/stac3092}, \href
  {https://ui.adsabs.harvard.edu/abs/2023MNRAS.518..724S} {518, 724}

\bibitem[\protect\citeauthoryear{{Spiniello} et~al.,}{{Spiniello}
  et~al.}{2021a}]{Spiniello2021a}
{Spiniello} C.,  et~al., 2021a, \mn@doi [\aap] {10.1051/0004-6361/202038936},
  \href {https://ui.adsabs.harvard.edu/abs/2021A&A...646A..28S} {646, A28}

\bibitem[\protect\citeauthoryear{{Spiniello} et~al.,}{{Spiniello}
  et~al.}{2021b}]{Spiniello2021b}
{Spiniello} C.,  et~al., 2021b, \mn@doi [\aap] {10.1051/0004-6361/202140856},
  \href {https://ui.adsabs.harvard.edu/abs/2021A&A...654A.136S} {654, A136}

\bibitem[\protect\citeauthoryear{{Stringer}, {Trujillo}, {Dalla Vecchia}  \&
  {Martinez-Valpuesta}}{{Stringer} et~al.}{2015}]{Stringer2015}
{Stringer} M.,  {Trujillo} I.,  {Dalla Vecchia} C.,   {Martinez-Valpuesta} I.,
  2015, \mn@doi [\mnras] {10.1093/mnras/stv455}, \href
  {https://ui.adsabs.harvard.edu/abs/2015MNRAS.449.2396S} {449, 2396}

\bibitem[\protect\citeauthoryear{{Tasca} et~al.,}{{Tasca}
  et~al.}{2009}]{Tasca2009}
{Tasca} L.~A.~M.,  et~al., 2009, \mn@doi [\aap] {10.1051/0004-6361/200912213},
  \href {https://ui.adsabs.harvard.edu/abs/2009A&A...503..379T} {503, 379}

\bibitem[\protect\citeauthoryear{{Thomas} et~al.,}{{Thomas}
  et~al.}{2019}]{Thomas2019}
{Thomas} R.,  et~al., 2019, \mn@doi [\aap] {10.1051/0004-6361/201935813}, \href
  {https://ui.adsabs.harvard.edu/abs/2019A&A...630A.145T} {630, A145}

\bibitem[\protect\citeauthoryear{{Tortora}, {Napolitano}, {Saglia},
  {Romanowsky}, {Covone}  \& {Capaccioli}}{{Tortora}
  et~al.}{2014}]{Tortora2014}
{Tortora} C.,  {Napolitano} N.~R.,  {Saglia} R.~P.,  {Romanowsky} A.~J.,
  {Covone} G.,   {Capaccioli} M.,  2014, \mn@doi [\mnras]
  {10.1093/mnras/stu1712}, \href
  {https://ui.adsabs.harvard.edu/abs/2014MNRAS.445..162T} {445, 162}

\bibitem[\protect\citeauthoryear{{Tortora} et~al.,}{{Tortora}
  et~al.}{2016}]{Tortora2016}
{Tortora} C.,  et~al., 2016, \mn@doi [\mnras] {10.1093/mnras/stw184}, \href
  {https://ui.adsabs.harvard.edu/abs/2016MNRAS.457.2845T} {457, 2845}

\bibitem[\protect\citeauthoryear{{Tortora} et~al.,}{{Tortora}
  et~al.}{2018}]{Tortora2018}
{Tortora} C.,  et~al., 2018, \mn@doi [\mnras] {10.1093/mnras/sty2564}, \href
  {https://ui.adsabs.harvard.edu/abs/2018MNRAS.481.4728T} {481, 4728}

\bibitem[\protect\citeauthoryear{{Tortora} et~al.,}{{Tortora}
  et~al.}{2020}]{Tortora2020}
{Tortora} C.,  et~al., 2020, \mn@doi [\aap] {10.1051/0004-6361/202038373},
  \href {https://ui.adsabs.harvard.edu/abs/2020A&A...638L..11T} {638, L11}

\bibitem[\protect\citeauthoryear{{Trujillo}, {Conselice}, {Bundy}, {Cooper},
  {Eisenhardt}  \& {Ellis}}{{Trujillo} et~al.}{2007}]{Trujillo2007}
{Trujillo} I.,  {Conselice} C.~J.,  {Bundy} K.,  {Cooper} M.~C.,  {Eisenhardt}
  P.,   {Ellis} R.~S.,  2007, \mn@doi [\mnras]
  {10.1111/j.1365-2966.2007.12388.x}, \href
  {https://ui.adsabs.harvard.edu/abs/2007MNRAS.382..109T} {382, 109}

\bibitem[\protect\citeauthoryear{{Turner} et~al.,}{{Turner}
  et~al.}{2021}]{Turner2021}
{Turner} S.,  et~al., 2021, \mn@doi [\mnras] {10.1093/mnras/stab653}, \href
  {https://ui.adsabs.harvard.edu/abs/2021MNRAS.503.3010T} {503, 3010}

\bibitem[\protect\citeauthoryear{{Valentinuzzi} et~al.,}{{Valentinuzzi}
  et~al.}{2010a}]{Valentinuzzi2010b}
{Valentinuzzi} T.,  et~al., 2010a, \mn@doi [\apj]
  {10.1088/0004-637X/712/1/226}, \href
  {https://ui.adsabs.harvard.edu/abs/2010ApJ...712..226V} {712, 226}

\bibitem[\protect\citeauthoryear{{Valentinuzzi} et~al.,}{{Valentinuzzi}
  et~al.}{2010b}]{Valentinuzzi2010a}
{Valentinuzzi} T.,  et~al., 2010b, \mn@doi [\apjl]
  {10.1088/2041-8205/721/1/L19}, \href
  {https://ui.adsabs.harvard.edu/abs/2010ApJ...721L..19V} {721, L19}

\bibitem[\protect\citeauthoryear{{Vietri} et~al.,}{{Vietri}
  et~al.}{2022}]{Vietri2021}
{Vietri} G.,  et~al., 2022, \mn@doi [\aap] {10.1051/0004-6361/202141072}, \href
  {https://ui.adsabs.harvard.edu/abs/2022A&A...659A.129V} {659, A129}

\bibitem[\protect\citeauthoryear{{Wellons} et~al.,}{{Wellons}
  et~al.}{2015}]{Wellons2015}
{Wellons} S.,  et~al., 2015, \mn@doi [\mnras] {10.1093/mnras/stv303}, \href
  {https://ui.adsabs.harvard.edu/abs/2015MNRAS.449..361W} {449, 361}

\bibitem[\protect\citeauthoryear{{Wellons} et~al.,}{{Wellons}
  et~al.}{2016}]{Wellons2016}
{Wellons} S.,  et~al., 2016, \mn@doi [\mnras] {10.1093/mnras/stv2738}, \href
  {https://ui.adsabs.harvard.edu/abs/2016MNRAS.456.1030W} {456, 1030}

\bibitem[\protect\citeauthoryear{{Wright} et~al.,}{{Wright}
  et~al.}{2010}]{Wright2010}
{Wright} E.~L.,  et~al., 2010, \mn@doi [\aj] {10.1088/0004-6256/140/6/1868},
  \href {https://ui.adsabs.harvard.edu/abs/2010AJ....140.1868W} {140, 1868}

\bibitem[\protect\citeauthoryear{{Wright}, {Lagos}, {Davies}, {Power},
  {Trayford}  \& {Wong}}{{Wright} et~al.}{2019}]{Wright2019}
{Wright} R.~J.,  {Lagos} C. d.~P.,  {Davies} L. J.~M.,  {Power} C.,  {Trayford}
  J.~W.,   {Wong} O.~I.,  2019, \mn@doi [\mnras] {10.1093/mnras/stz1410}, \href
  {https://ui.adsabs.harvard.edu/abs/2019MNRAS.487.3740W} {487, 3740}

\bibitem[\protect\citeauthoryear{{Wu} et~al.,}{{Wu} et~al.}{2020}]{Wu2020}
{Wu} P.-F.,  et~al., 2020, \mn@doi [\apj] {10.3847/1538-4357/ab5fd9}, \href
  {https://ui.adsabs.harvard.edu/abs/2020ApJ...888...77W} {888, 77}

\bibitem[\protect\citeauthoryear{{Yoon}, {Im}  \& {Kim}}{{Yoon}
  et~al.}{2017}]{Yoon2017}
{Yoon} Y.,  {Im} M.,   {Kim} J.-W.,  2017, \mn@doi [\apj]
  {10.3847/1538-4357/834/1/73}, \href
  {https://ui.adsabs.harvard.edu/abs/2017ApJ...834...73Y} {834, 73}

\bibitem[\protect\citeauthoryear{{Zanisi}, {Shankar}, {Bernardi}, {Mei}  \&
  {Huertas-Company}}{{Zanisi} et~al.}{2021}]{zanisi21}
{Zanisi} L.,  {Shankar} F.,  {Bernardi} M.,  {Mei} S.,   {Huertas-Company} M.,
  2021, \mn@doi [\mnras] {10.1093/mnrasl/slab056}, \href
  {https://ui.adsabs.harvard.edu/abs/2021MNRAS.505L..84Z} {505, L84}

\bibitem[\protect\citeauthoryear{{van Dokkum} et~al.,}{{van Dokkum}
  et~al.}{2010}]{vanDokkum2010}
{van Dokkum} P.~G.,  et~al., 2010, \mn@doi [\apj]
  {10.1088/0004-637X/709/2/1018}, \href
  {https://ui.adsabs.harvard.edu/abs/2010ApJ...709.1018V} {709, 1018}

\bibitem[\protect\citeauthoryear{{van der Wel} et~al.,}{{van der Wel}
  et~al.}{2012}]{vanderWel2012}
{van der Wel} A.,  et~al., 2012, \mn@doi [\apjs] {10.1088/0067-0049/203/2/24},
  \href {https://ui.adsabs.harvard.edu/abs/2012ApJS..203...24V} {203, 24}

\bibitem[\protect\citeauthoryear{{van der Wel} et~al.,}{{van der Wel}
  et~al.}{2014}]{vanderWel2014}
{van der Wel} A.,  et~al., 2014, \mn@doi [\apj] {10.1088/0004-637X/788/1/28},
  \href {https://ui.adsabs.harvard.edu/abs/2014ApJ...788...28V} {788, 28}

\makeatother
\end{thebibliography}


\appendix

\section{Size measurements of \textit{red nugget sample}}\label{app:size_validation} 

The selection of the {\it red nugget sample} described by \citetalias{Lisiecki2022} is based on the size measurements performed by \citealt{Krywult2017}. At higher redshift ($\rm{z>0.2}$),
the ground-based image resolution limits recovering accurate structural parameters. Moreover, in dense regions of the survey, the target galaxy's outskirts can overlap with the outskirts of their companions. 
It may limit the fraction of small/compact galaxy light profiles that can be recovered/measured securely. In this case, the galaxy light profiles should be fitted simultaneously to all objects ~\citep[e.g.][]{vanderWel2012,Mowla2019}. 

 \begin{figure}
	\centerline{\includegraphics[width=0.49\textwidth]{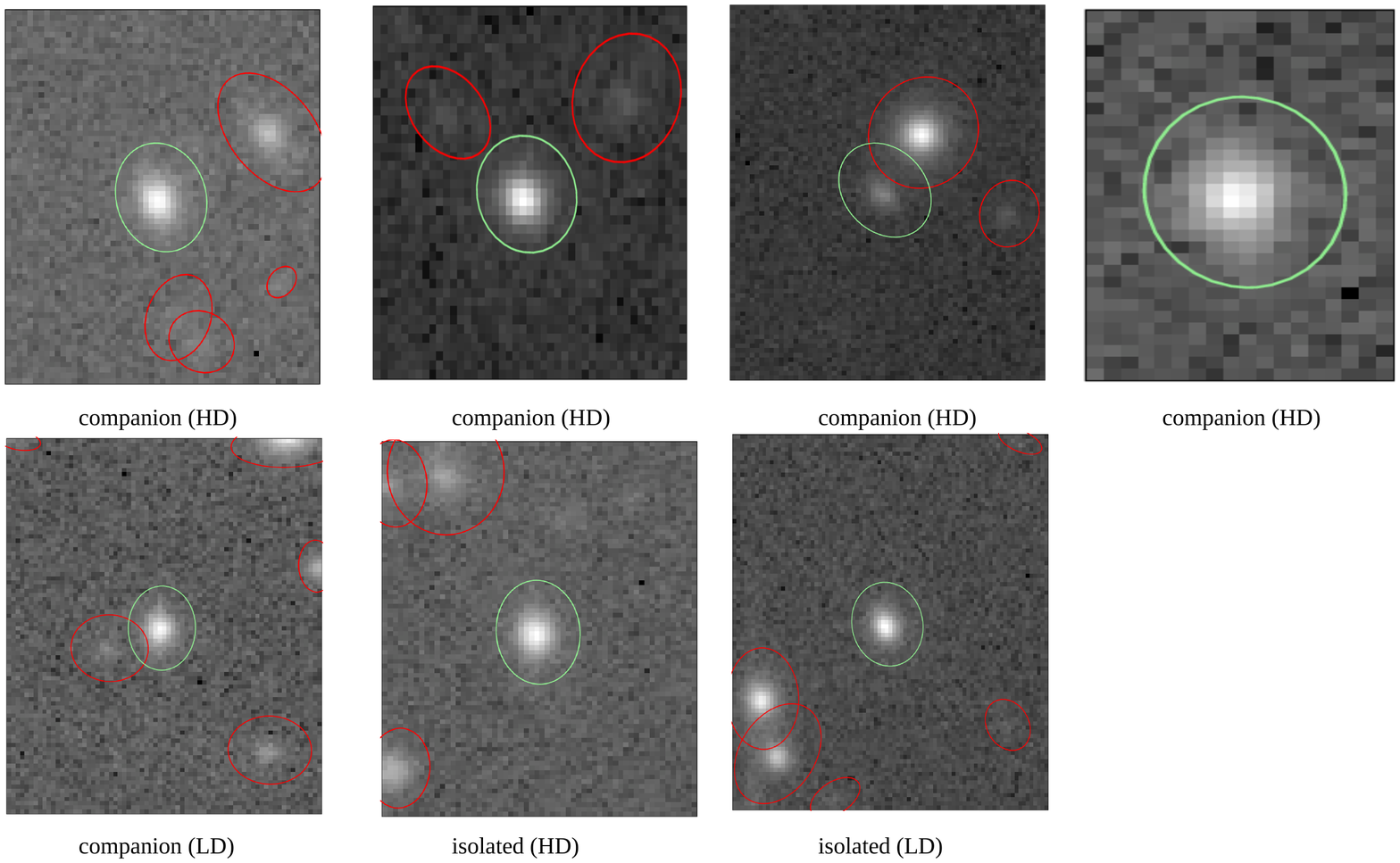}}
	\caption{Examples of postage stamps of target galaxies (in green ellipses) and other
galaxies (red ellipses) in the area 4 times larger than the size of the target galaxy. The top row consists of the stamps of galaxies with a companion found in the HD environment (for the galaxy shown in the last stamp, only the small size stamp is preserved without revealing companions), and the bottom row consists of examples of isolated galaxies in LD and HD environments. }
	\label{fig:apertury}
\end{figure} 

\begin{figure}
	\centerline{\includegraphics[width=0.49\textwidth]{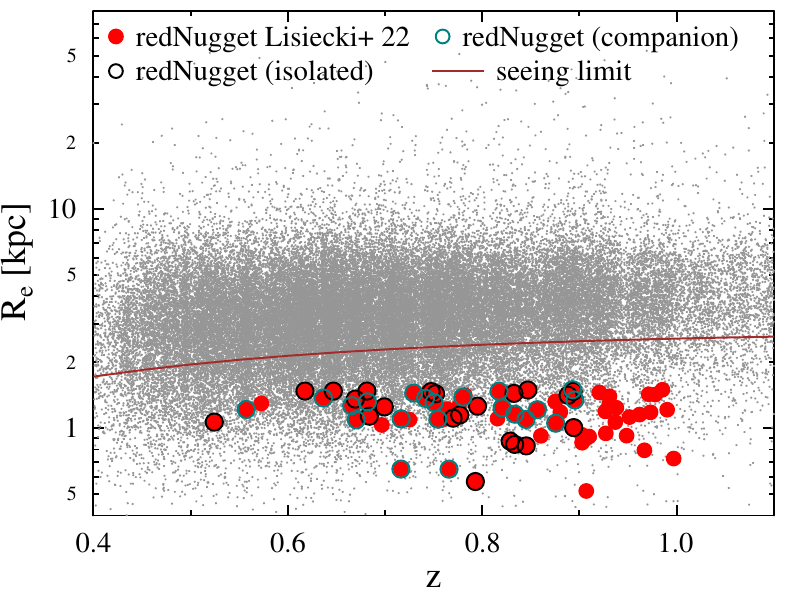}}
	\caption{{Seeing-limit plot for VIPERS galaxies. The \textit{red nugget sample} is {under} the seeing limit independent of whether it is an isolated galaxy, i.e. a single galaxy
on the stamp, or its outskirts overlap with the outer parts of the companion.}}
	\label{fig:seeing_limit}
\end{figure}

We find that half of the {\it red nugget sample} (as well as of the complete red nugget sample from \citetalias{Lisiecki2022}) have outer regions close to the outer regions of another galaxy, see examples in Fig.~\ref{fig:apertury}. The green and red elliptical apertures in the postage stamps enclose the \textit{red nugget} and neighbouring galaxy, respectively. 
The aperture size is 2.5 times the elliptical Kron radius. The procedure applied by \citealt{Krywult2017} simultaneously fits the S\'ersic profile by GALFIT~\citep{Peng2002} to the target galaxy and the galaxies in their neighbourhood when the aperture ellipse of an accompanying object increased by a factor of 1.5 overlaps with that of the main target. 
Such an approach allows us to recover structural measurements in the densest regions of the survey's ground-based (seeing-limited) images (see Fig. 2 in~\citetalias{Lisiecki2022} and Fig.~\ref{fig:seeing_limit}).

\begin{figure*}
  \centering
{\includegraphics[width=0.99\linewidth]{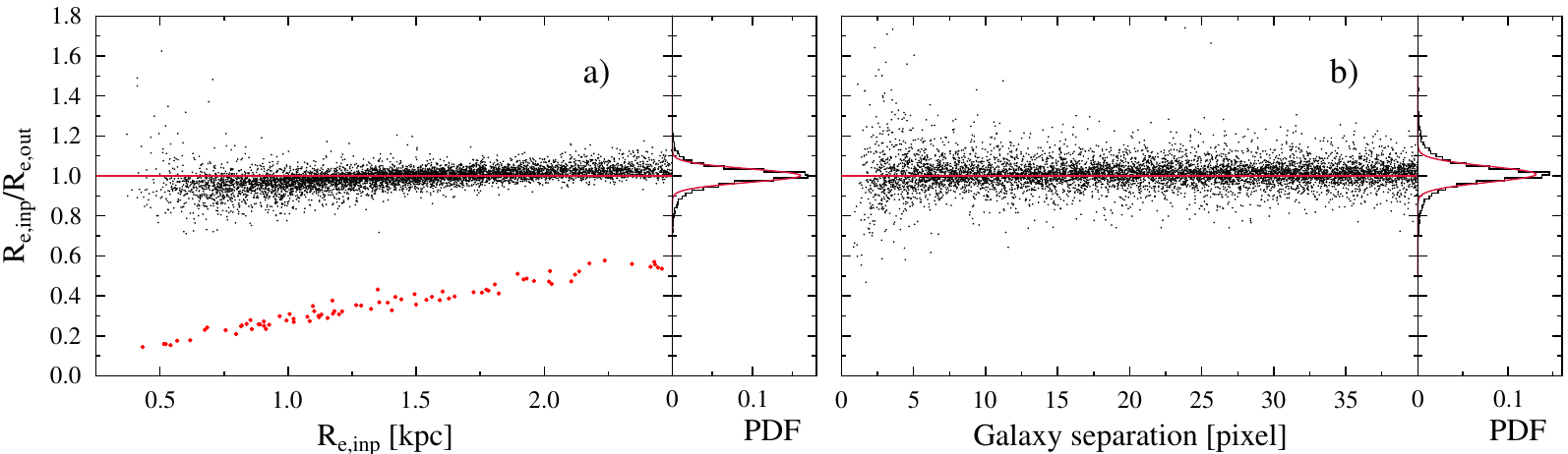}}
  \caption{{ Monte Carlo results of the effective radius recovering tests for the
set of simulated galaxies: left panel: impact of the effective radius $R_{\rm e,out}$
fit as a function of the model $R_{\rm e,inp}$, with (black dots) PSF
deconvolution and without it (red dots); right panel: impact of neighbouring galaxies
on the effective radius $R_{\rm e,out}$ fit as a function of galaxy separation. 
In both panels, the probability distribution function (PDF) of $R_{\rm e,inp}$/$R_{\rm e,out}$ corresponding to each test is presented by the black line, whereas the red line shows the Gaussian fit. 
 }   }
  \label{fig:re_simulations}
\end{figure*}

{To precisely verify the accuracy of the effective radii of the investigated {\it red nugget} galaxies
obtained from ground-based CCD frames, we perform a set of the Monte Carlo simulations. 
The Monte Carlo simulation was based on 10,000 artificial galaxies in the redshift
range $0.5\le z\le0.9$. The artificial galaxy images were generated  using the
single component S\'ersic profile with the randomly selected parameters' values;
semi-major axis: $0.5\le a_e\le 3$ [pixel], S\'ersic index: $0.5\le n\le5$, 
magnitude: $20.5\le i{\rm-band\ mag}\le 22.5$, axis ratio: $0.2\le b/a\le 0.9$
and position angle: $0^\circ\le {\rm PA} \le 180^\circ$. Each galaxy image was convolved with the $i$-band point spread function (PSF) of the CFHTLS (\citealt{Krywult2017}). 
The PSF was computed at the positions of randomly selected 6000 VIPERS galaxies on the CFHTLS tiles. 
Then, the artificial galaxy was added to the centre
of the background image, $80\times80$ pixels size each, obtained by mosaicing
different portions of the object-free regions of the CFHTLS. 
The S\'ersic profile parameters were fitted by the GALFIT. 
The results are shown in Fig.~\ref{fig:re_simulations}.
}


{
The left panel of Fig.~\ref{fig:re_simulations} presents the ratio of the model to the fitted radius, $R_{\rm e,inp}/R_{\rm e,out}$, versus the model $R_{\rm e,inp}$ value. 
For comparison, a small subsample of the model galaxies was fitted by GALFIT without PSF
deconvolution (marked with red dots), which shows a significant demotion of the recovered half-light
radius value and indicate that the deconvolution in the fitting procedure is
essential to obtain reliable results, as shown with the black dots. With
the deconvolution, these points are almost horizontally distributed with only a
little, about 10\%, over-determined recovered $R_{\rm e,out}$ value for the
smallest galaxies, for which the effective radius is smaller than 1~kpc, less than
the lower limit of our {\it red nugget} sample. The  histogram presented on the 
right plot side of the panel shows narrow $R_{\rm e,inp}/R_{\rm e,out}$
distribution with the Gaussian dispersion $1\sigma$ equal to 0.03.
}

{
From the presented test, we conclude that PSF deconvolution applied during the
GALFIT galaxy profile fitting  significantly improves the accuracy of the
effective radius reconstruction. We found that even if the galaxies are located
below the seeing limit, see Fig.~\ref{fig:re_simulations}, it is possible to reconstruct their effective radius correctly.
}

{In the next step, we test the influence of the nearby second galaxy on the
investigated galaxy's recovered effective radius $\rm{R_e}$. Using the same procedure
as described above, the neighbourhood galaxy image was modelled and added to the
postage stamp image with the previously generated target galaxy. The separation
between both galaxy centres was generated randomly in the range from $\rm{r_{e,c}}$ to
40 pixels, where $\rm{r_{e,c}}$ is the circularized  half-light radius of the target
galaxy. The right panel, b), of Fig.~\ref{fig:re_simulations} shows that in the wide range of the
galaxy centre separation, the target galaxy $R_{\rm e,out}$ fitted value is 
almost insensitive to the presence of the neighbour location. The presented
histogram of $R_{\rm e,inp}/R_{\rm e,out}$ is narrow with $1\sigma=0.04$.
However, for the galaxy centre separation of less than about 5 pixels, the
presence of another galaxy decreases the accuracy of the recovered $R_{\rm
e,out}$. This critical distance is significantly lower than the real minimal
separation between the {\it red nugget} and nearest neighbour galaxy, which is
equal to 12 pixels, see Fig.~\ref{fig:re_simulations}.
}

{
The Monte Carlo tests presented above lead us to the conclusion that the coming
from seeing limited ground-based CFHTLS observation {\it red nugget} effective
radii are correctly recovered.}

\begin{figure}
	\centerline{\includegraphics[width=0.49\textwidth]{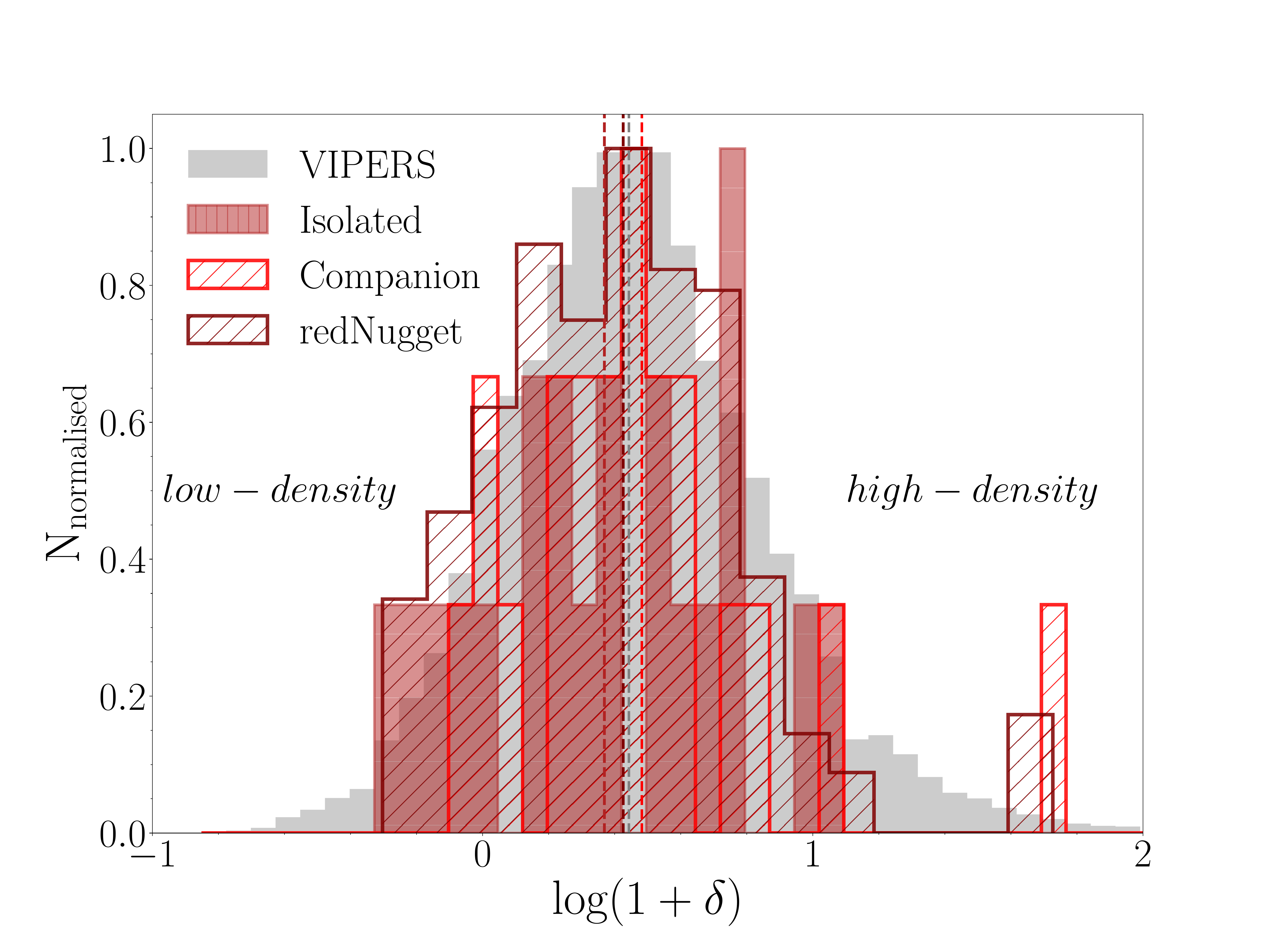}}
	\caption{{Distribution of $\rm{log(1+\delta)}$ for \textit{red nugget sample}  (in brown) separated into the ones that are isolated, i.e. single galaxy on the stamp (in firebrick) and the ones whose outskirts overlap with the outskirts of companions (in red).}}
	\label{fig:density_distribution_companion}
\end{figure}

Moreover, finding that for half  of the {\it red nugget sample} their outskirts overlap with the outskirts of companions suggests that our sample does not suffer from incompleteness issues. However, there is no dependence of the number of {\it red nuggets} with companions on the environment: {\it red nuggets} whose outskirts overlap with the outskirts of other galaxies are found in LD environments (4 galaxies) as well as in the HD environments (4 galaxies). Red nuggets with companions are also found in the $\delta$-end of the distribution of VIPERS/red nugget sample (see Fig.~\ref{fig:density_distribution_companion}). 
Additionally, \citealt{Lisiecki2022} have shown that the number density of red nuggets stays in agreement with the other studies (see Fig.8 in~\citetalias{Lisiecki2022}).

\section{Colours of red nuggets and control samples}\label{app:colours} 

\begin{figure}
	\centerline{\includegraphics[width=0.49\textwidth]{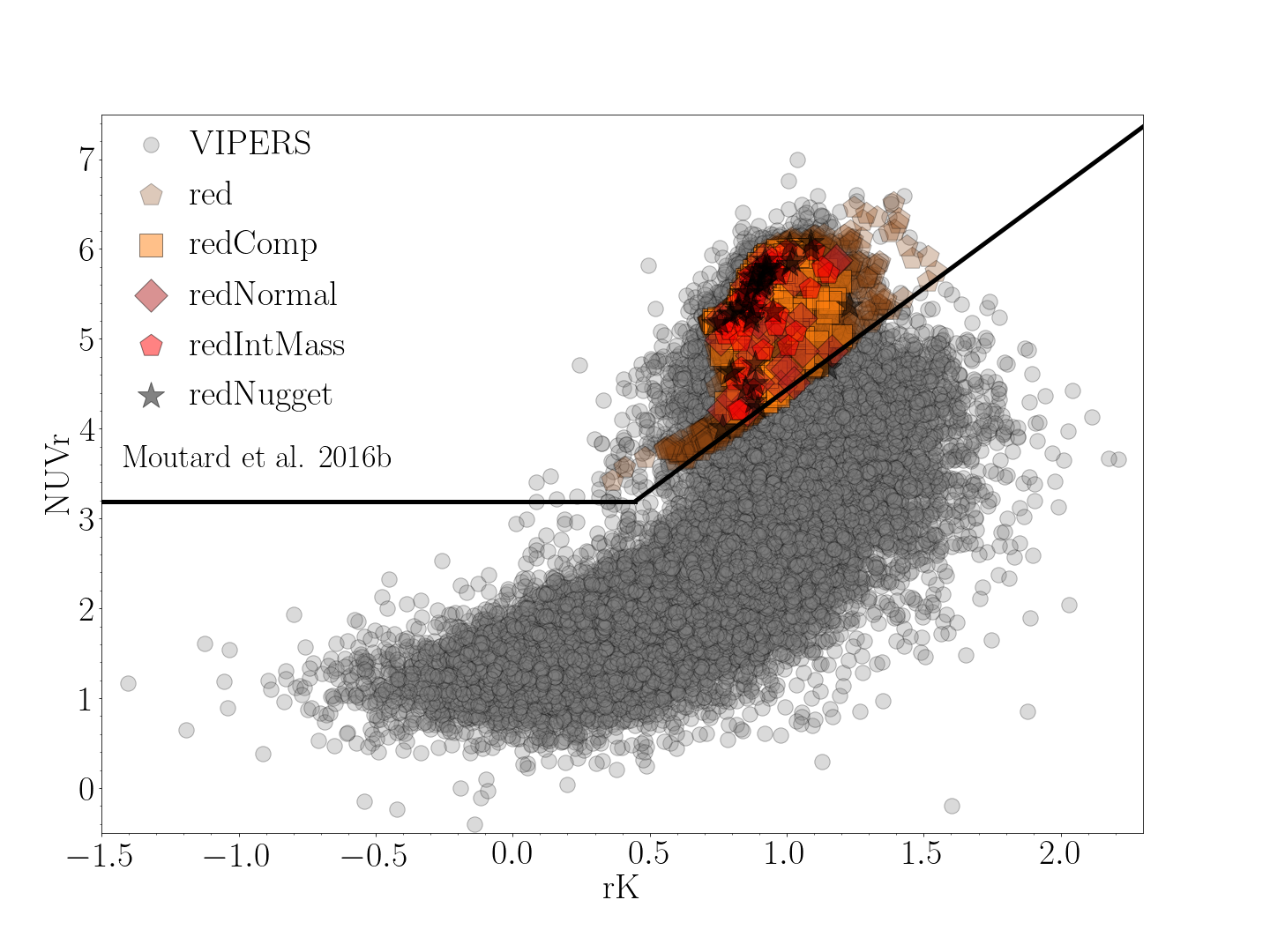}}
	\caption{NUVrK diagram of VIPERS galaxies (in gray) and \textit{red nuggets} (black stars) as well as control galaxies. Galaxies classified as red are separated from the blue ones by the solid black line as proposed by~\citealt{moutard2016b}. }
	\label{fig:sample_nuvrk}
\end{figure}

This work uses a mass-complete sample of VIPERS galaxies (see Sec.~\ref{sec:sample_selection}). 
The mass cuts were implemented in three redshift bins ($\rm{0.50<z\leq0.65}$, $\rm{0.65<z\leq0.85}$, $\rm{0.80<z\leq0.90}$) for passive and active galaxies~\citep{Davidzon2016}. 
To separate the \textit{red sample} we applied the cut in the NUVrK diagram proposed by~\cite{Arnouts2013, moutard2016b}. 
The classification of passive and active galaxies is based on the NUVrK diagram with a cut evolving with redshift~\citep[see Eq. 4 in][]{moutard2016b}.
In Fig.~\ref{fig:sample_nuvrk} the cut to separate red and blue galaxies at $\rm{z=0.7}$ is shown, but we applied cuts separately in each redshift bin. 
The selected red mass-complete sample (\textit{red sample}) is used as a control sample to represent the full diversity of red galaxies. 

We further evaluated the colours of our control samples using the NUVrK diagram (see Fig~\ref{fig:sample_nuvrk}). 
Control samples recreate the standard colour-colour division~\citep{moutard2016b}, placing each sample in the red cloud. 
While \textit{red normal-size}, \textit{red intermediate-mass} {and \textit{red compact-size}} samples occupy the same region of the NUVrK as \textit{red nuggets},  the \textit{red sample} is spanning over the entire red NUVrK region.

\section{Stellar ages of \textit{red nugget sample}}\label{app:stage_validation}

{
Stellar ages of the \textit{red nugget} sample are derived based on the synthetic stellar age-D4000 relation (BC03 models; see Fig.~\ref{fig:BC03_ages} and Sec.~\ref{sec:rednugget_ages}). The estimated stellar ages and the age of the Universe (we assume that the age of the galaxy is similar to the age of the Universe) at a given redshift are reported in Tab.~\ref{table:stellar_ages}. We note that we do not limit the stellar ages estimations so that the best fit can be found with the age exceeding the age of the galaxy at the given redshift.  }

	\begin{table*}
		\centering                         
		\begin{tabular}{P{1.5cm}  P{1.cm} P{1.cm} P{1.5cm} }    
			\hline 
			VIPERS & z  & $\rm{t_z}$ & Age [Gyr]\\
			\hline 
            \hline
102164621 & 0.84 & 6.79 & 2.93 \\ 
103137838 & 0.75 & 7.27 & 4.92 \\ 
104169502 & 0.77 & 7.16 & 1.75 \\ 
104243009 & 0.74 & 7.32 & 5.20 \\ 
105149946 & 0.78 & 7.10 & 4.10 \\ 
107109094 & 0.82 & 6.89 & 1.79 \\ 
107156787 & 0.68 & 7.68 & 2.14 \\ 
109135372 & 0.83 & 6.84 & 3.99 \\ 
109179636 & 0.52 & 8.74 & 2.74 \\ 
111088775 & 0.82 & 6.89 & 3.12 \\ 
111091579 & 0.67 & 7.74 & 1.66 \\ 
111126369 & 0.86 & 6.69 & 3.68 \\ 
112184594 & 0.75 & 7.27 & 5.54 \\ 
112195565 & 0.67 & 7.74 & 8.13 \\ 
115108461 & 0.83 & 6.84 & 8.46 \\ 
121089481 & 0.78 & 7.10 & 3.72 \\ 
122030630 & 0.68 & 7.68 & 2.70 \\ 
122057921 & 0.70 & 7.56 & 3.61 \\ 
123089330 & 0.73 & 7.38 & 4.04 \\ 
124039182 & 0.83 & 6.84 & 1.72 \\ 
124039429 & 0.85 & 6.74 & 2.98 \\ 
124070857 & 0.89 & 6.54 & 3.20 \\ 
401174682 & 0.67 & 7.74 & 2.46 \\ 
401213761 & 0.85 & 6.74 & 1.93 \\ 
402071577 & 0.56 & 8.45 & 3.57 \\ 
402075551 & 0.88 & 6.59 & 1.93 \\ 
402173836 & 0.77 & 7.16 & 2.09 \\ 
403027664 & 0.68 & 7.68 & 2.81 \\ 
403037148 & 0.89 & 6.54 & 6.79 \\ 
403054648 & 0.89 & 6.54 & 1.51 \\ 
403058809 & 0.89 & 6.54 & 5.12 \\ 
403086972 & 0.72 & 7.44 & 0.61 \\ 
403099223 & 0.75 & 7.27 & 0.92 \\ 
403101616 & 0.90 & 6.50 & 2.98 \\ 
404066979 & 0.80 & 7.00 & 5.12 \\ 
404139209 & 0.65 & 7.86 & 3.59 \\ 
408062025 & 0.72 & 7.44 & 8.87 \\ 
408062039 & 0.62 & 8.05 & 3.96 \\ 
409065277 & 0.79 & 7.05 & 1.65 \\ 
409068570 & 0.75 & 7.27 & 3.46 \\ 
410067404 & 0.64 & 7.92 & 1.95 \\ 
410072423 & 0.83 & 6.84 & 3.87 \\
			\hline
		\end{tabular}
		\caption{{Stellar ages derived from the synthetic stellar age-D4000 relation for the sample of \textit{red nugget sample}. The VIPERS ID, spectroscopic redshift, z, and the corresponding age of the Universe (approximately corresponding to the age of the galaxy, $\rm{t_z}$), at a given redshift, are reported.  } }             
		\label{table:stellar_ages}     
	\end{table*}


\bsp	
\label{lastpage}
\end{document}